\documentclass[
 reprint,
superscriptaddress,
 amsmath,amssymb,
 aps,
pra,
nofootinbib,longbibliography]{revtex4-1}
\usepackage{hyperref}
\newcommand{\bi}{\mathbf}
\usepackage{graphicx}
\usepackage{dcolumn}
\usepackage{caption}
\usepackage{subcaption}
\usepackage{bm}
\usepackage{verbatim}
\usepackage{float}
\usepackage{balance}
\usepackage{natbib}

\newcommand{\rme}{\mathrm{e}}
\newcommand{\me}{m_e}

\usepackage{color}
\newcommand{\Radd}[1]{\textcolor{red}{#1}}

\newcommand{\Heff}{\hat{H}_{\mathrm{eff}}}

\begin{document}


\title{Polaritonic Hofstadter Butterfly and Cavity-Control of the Quantized Hall Conductance} 

\author{Vasil~Rokaj}
\email{vasil.rokaj@cfa.harvard.edu}
\affiliation{Max Planck Institute for the Structure and Dynamics of Matter,
Center for Free Electron Laser Science, 22761 Hamburg, Germany}
\affiliation{ITAMP, Harvard-Smithsonian Center for Astrophysics, Cambridge, MA 02138, USA}

\author{Markus Penz}
\affiliation{Department of Mathematics, University of Innsbruck,
Technikerstraße 13/7, A-6020 Innsbruck, Austria}

\author{Michael A. Sentef}
\affiliation{Max Planck Institute for the Structure and Dynamics of Matter,
Center for Free Electron Laser Science, 22761 Hamburg, Germany}

\author{Michael~Ruggenthaler}

\affiliation{Max Planck Institute for the Structure and Dynamics of Matter,
Center for Free Electron Laser Science, 22761 Hamburg, Germany}

\author{Angel~Rubio}
\email{angel.rubio@mpsd.mpg.de}
\affiliation{Max Planck Institute for the Structure and Dynamics of Matter,
Center for Free Electron Laser Science, 22761 Hamburg, Germany}
\affiliation{Center for Computational Quantum Physics (CCQ), Flatiron Institute, 162 Fifth Avenue, New York NY 10010}


\date{\today}

\begin{abstract}
In a previous work [Phys.~Rev.~Lett.~123, 047202 (2019)] a translationally invariant framework called quantum-electrodynamical Bloch (QED-Bloch) theory was introduced for the description of periodic materials in homogeneous magnetic fields and strongly coupled to the quantized photon field in the optical limit. For such systems, we show that QED-Bloch theory predicts the existence of fractal polaritonic spectra as a function of the cavity coupling strength. In addition, for the energy spectrum as a function of the relative magnetic flux we find that a terahertz cavity can modify the standard Hofstadter butterfly. In the limit of no quantized photon field, QED-Bloch theory captures the well-known fractal spectrum of the Hofstadter butterfly and can be used for the description of 2D materials in strong magnetic fields, which are of great experimental interest. As a further application, we consider Landau levels under cavity confinement and show that the cavity alters the quantized Hall conductance and that the Hall plateaus are modified as $\sigma_{xy}=e^2\nu/h(1+\eta^2)$ by the light-matter coupling $\eta$. Most of the aforementioned phenomena should be experimentally accessible and corresponding implications are discussed.

\end{abstract}

\pacs{Valid PACS appear here}
\maketitle


\section{Introduction}

The study of two-dimensional systems perpendicular to a strong homogeneous magnetic field has given rise to a plethora of macroscopic quantum phenomena known as the quantum Hall effects~\cite{40QHE}. This fundamental branch of condensed matter physics was ignited by the discovery of the integer Hall effect in 1980~\cite{Klitzing}, where the macroscopic Hall conductance $\sigma_{xy}$ exhibits quantized plateaus whose value depend solely on the fundamental charge $e$, Planck's constant $h$, and the filling factor $\nu$ in the picture of non-interacting Landau levels, $\sigma_{xy}=e^2\nu/h$. In the four decades after the fundamental discovery of the integer Hall effect, a great number of related phenomena like the fractional quantum Hall effect~\cite{TsuifractionalQHE, Laughlingfractional}, the quantum spin Hall effect~\cite{QuantumSpinHall}, the quantum anomalous Hall effect~\cite{anomalousHalleffect} and more recently the light-induced anomalous Hall effect~\cite{LightHalleffect} have been observed and studied theoretically, as they provide a unique platform to address strongly correlated electronic phases, topology  and strong light-matter phenomena. All these exciting developments have been reviewed beautifully in an article celebrating the 40 years anniversary of the quantum Hall effect~\cite{40QHE}.

In this cornucopia of phenomena offered by the quantum Hall setting there is another important phenomenon which stands out due to its fundamental nature, aesthetic beauty, and the connections that it provides between different branches of mathematics and physics, namely the Hofstadter butterfly~\cite{Hofstadter}. The Hofstadter butterfly is a fractal pattern that describes the energies of electrons on a periodic lattice perpendicular to a homogeneous magnetic field as a function of the relative magnetic flux $\Phi/\Phi_0$, where $\Phi$ is the magnetic flux through the fundamental unit cell of the lattice and $\Phi_0=h/e$ is the magnetic flux quantum. It was predicted by Hofstadter in 1976~\cite{Hofstadter} and it was proven to be a fractal by Avila and Jitomirskaya~\cite{Avila2006}. In recent years, due to the advent of Moir\'{e} materials~\cite{Moiremarvels, RubioMoire}, it has become possible to probe it experimentally, and signatures of the fractal spectrum have been observed in the magnetotransport properties of Moir\'{e} systems~\cite{DeanButterfly, WangButterfly, ForsytheButterfly}. Further, the physics of the Hofstadter butterfly has also been realized with ultracold atoms in optical lattices~\cite{ButterflyAidelsburger, ButterflyKetterle}.

Another pillar of modern quantum physics is quantum electrodynamics (QED), which describes the interaction of charged particles with photons~\cite{cohen1997photons, Weinberg, spohn2004}. In the last decade there has been a great interest in the regime of strong and ultrastrong light-matter interactions~\cite{kockum2019ultrastrong}, where light and matter lose their individual character and form hybrid quasiparticles known as polaritons~\cite{PolaritonPanorama}. Many different platforms and routes have been explored to reach strong light-matter coupling and several unprecedented phenomena involving polaritonic states have been observed. Modifications of chemical properties and chemical reactions have been achieved through coupling to vacuum fields in polaritonic chemistry~\cite{ebbesen2016, hutchison2013, hutchison2012, orgiu2015, feist2017polaritonic, galego2016, flick2017, schafer2019modification}. Cavity control of excitons has been studied~\cite{LatiniRonca, ExcitonControl} and exciton-polariton condensation has been achieved~\cite{kasprzak2006, KeelingKenaCohen}. It has been suggested that light-matter interactions modify the electron-phonon coupling and the critical temperature of superconductors~\cite{SchlawinSuperconductivity, AtacSuperconductivity, sentef2018, Galitski}, with the first experimental evidence already at hand~\cite{A.Thomas2019}. Further, the implications of coupling to chiral electromagnetic fields is currently investigated~\cite{ChiralCavities, ChiralPetersen67, PRAChiral, ChiralQuantumOptics, SentefRonca}, and the possibility of cavity-induced ferroelectric phases has been proposed~\cite{LatiniFerro, Demlerferro}.

As a synthesis of QED and the quantum Hall setting, quantum Hall systems under cavity confinement have been studied both experimentally and theoretically, in the integer~\cite{Hagenmuller2010cyclotron, rokaj2019, Keller2020, ScalariScience, li2018} and the correlated fractional~\cite{AtacPRL, SmolkaAtac} regime, and ultrastrong coupling to the photon field and modifications of their transport properties~\cite{paravacini2019} have been demonstrated. Recently, a theoretical mechanism for a cavity-mediated hopping in the integer regime was also proposed~\cite{CiutiHopping} and the breakdown of the topological protection of the integer Hall effect due to cavity vacuum fields was demonstrated experimentally~\cite{FaistCavityHall}.

\begin{figure}[H]
\begin{center}
  \includegraphics[width=0.7\linewidth]{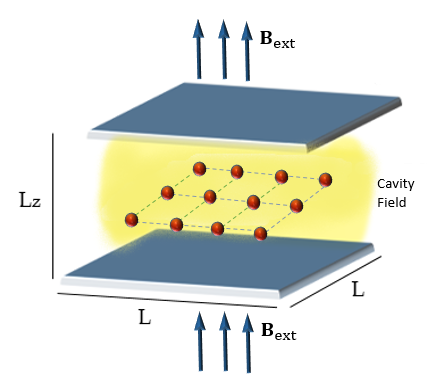}
\caption{Cartoon depiction of a two-dimensional periodic material confined inside a cavity with mirrors of length $L$ and area $S=L^2$. The distance between the cavity mirrors is $L_z$. The whole system is placed perpendicular to a classical homogeneous magnetic field $\bi{B}_{\textrm{ext}}$. We note that typically in experimental setups the space between the 2D material and the cavity is filled with a dielectric medium.}  
\label{2D Solid Cavity}
\end{center}
\end{figure}

In this article, we focus on this emerging field of quantum Hall systems strongly coupled to the quantized photon-field originating from a cavity (see Fig.~\ref{2D Solid Cavity}). To be more precise, we investigate the modification of two particular phenomena due to strong vacuum fluctuations inside a cavity: (i) the Hofstadter butterfly and (ii) the quantization of the Hall conductance in the integer regime. To describe these quantum Hall systems and phenomena in the cavity setting we employ the recently-introduced quantum-electrodynamical Bloch (QED-Bloch) theory~\cite{rokaj2019, RokajThesis}. QED-Bloch theory solves the long-standing problem of broken translational invariance due to an external magnetic field and provides a translationally symmetric framework for periodic systems in the presence of a homogeneous magnetic field and strongly coupled to the quantized photon field in the optical limit (or dipole approximation). Our main findings are:
\begin{itemize}
   \item\textbf{Polaritonic Hofstadter Butterfly.}---For a 2D periodic system perpendicular to a homogeneous magnetic field and under cavity confinement we find that for the energy spectrum a self-similar pattern emerges as a function of the light-matter coupling $\eta$ as depicted in Fig.~\ref{Fractal_pol_g5}. This is an extension of the standard Hofstadter butterfly~\cite{Hofstadter} to the polaritonic (light-matter) setting of cavity QED which introduces the concept of polaritonic fractal spectra. We call this phenomenon the polaritonic Hofstadter butterfly. In addition, we compute the energy spectrum for a periodic material as a function of the relative magnetic flux and find that for a terahertz cavity~\cite{ScalariScience, paravacini2019, li2018} the standard Hofstadter butterfly gets modified due to the strong vacuum fluctuations of the photon field (see Fig.~\ref{Hofstadter fluctuations large}).
   
   \item\textbf{Periodic Materials in Homogeneous Magnetic Fields.}---In the semi-classical limit of no quantized field our QED-Bloch theory recovers the standard results of condensed matter systems in strong magnetic fields like the Hofstadter butterfly~\cite{Hofstadter}, the Landau levels~\cite{Landau} and the quantization of the Hall conductance~\cite{Klitzing, LaughlinPRB} and provides a first-principles framework for the description of periodic materials in strong magnetic fields. In addition, a dual relation between the minimal-coupling Hamiltonian and the tight-binding models with the Peierls substitution is established (see Fig.~\ref{Butterfly Duality}).
   
   \item\textbf{Cavity Modification of the Integer Hall Effect.}---For a 2D electron gas consisting of Landau levels strongly coupled to the cavity field we find that the quantized Hall conductance gets modified. The Hall plateaus inside the cavity depend on the light-matter coupling constant $\eta$ as $\sigma_{xy}=e^2\nu/h(1+\eta^2)$. This modification is a consequence of the formation of hybrid quasiparticle states between the Landau levels and the cavity photons known as Landau polaritons~\cite{ScalariScience, paravacini2019, rokaj2019, Hagenmuller2010cyclotron}.  
\end{itemize}

\textit{Outline of the Article.}---In Section~\ref{QED Bloch Theory} we recapitulate the basic steps of the construction of QED-Bloch theory and the restoration of translational symmetry. In Section~\ref{QED Bloch Expansion} we construct the QED-Bloch ansatz which we use for the description of periodic materials in strong magnetic fields and strongly coupled to the quantized photon field and we derive the QED-Bloch central equation. In Section~\ref{Polaritonic Butterfly} we show that for periodic materials in strong magnetic fields and under cavity confinement there exists a polaritonic fractal spectrum as a function of the light-matter coupling constant. In Section~\ref{Harper Duality} we demonstrate that there is a duality between the minimal-coupling Hamiltonian and the tight-binding models with the Peierls phase. In Section~\ref{Modification Hall Conductance} we compute the Hall conductance for non-interacting Landau levels coupled to the cavity field and we show that the cavity modifies the quantization of the Hall conductance. Finally, in Section~\ref{Conclusions} we conclude and highlight the future perspectives of this work. 

\section{QED-Bloch Theory}\label{QED Bloch Theory}

The description of periodic materials in the presence of a homogeneous magnetic field has been a long-standing problem for condensed matter physics. The problem arises due to the fact that, although the magnetic field is homogeneous throughout the whole material, in the minimally-coupled Schr\"{o}dinger equation the electrons couple to the spatially inhomogeneous vector potential of the electromagnetic field. Thus, translational symmetry is broken and Bloch theory is not applicable.

Recently, a translationally invariant, quantum-electro\-dynamical framework for the description of periodic solids in homogeneous magnetic fields was introduced by the authors~\cite{rokaj2019}. Due to the fact that in this framework translational symmetry is restored in the higher-dimensional configuration space of both electrons and photons, in which Bloch's theorem can still be applied, it was named quantum-electrodynamical Bloch (QED-Bloch) theory. Before we proceed with the application of this framework we would like to briefly recapitulate the basic steps in the construction of QED-Bloch theory. 

Our starting point is the Pauli-Fierz Hamiltonian for $N$ interacting electrons in a periodic potential and in the presence of a classical, homogeneous magnetic field and further coupled to the quantized photon field~\cite{cohen1997photons, spohn2004, rokaj2017},
\begin{eqnarray}\label{velgauge} 
\hat{H}&=&\sum\limits^{N}_{j=1}\left[\frac{1}{2\me}\left(\mathrm{i}\hbar \mathbf{\nabla}_{j}+e\hat{\mathbf{A}}(\bi{r}_j)+e\mathbf{A}_{\textrm{ext}}(\bi{r}_j)\right)^2 +v_{\textrm{ext}}(\mathbf{r}_{j})\right]\nonumber\\
&+&\frac{1}{4\pi\epsilon_0}\sum\limits^{N}_{j< k}\frac{e^2}{|\mathbf{r}_j-\mathbf{r}_k|}+\sum\limits_{\bm{\kappa},\lambda}\hbar\omega(\bm{\kappa})\left[\hat{a}^{\dagger}_{\bm{\kappa},\lambda}\hat{a}_{\bm{\kappa},\lambda}+\frac{1}{2}\right],
\end{eqnarray}
where we neglected the Pauli (Stern-Gerlach) term  $\hat{\bm{\sigma}}\cdot\hat{ \bi{B}}(\bi{r})$ as it is commonly done for the description of the Hofstadter butterfly~\cite{Hofstadter} and the Landau levels in the quantum Hall effect~\cite{LaughlinPRB, Landau}. Here $\mathbf{A}_{\textrm{ext}}(\mathbf{r})$ is the external vector potential which gives rise to the homogeneous magnetic field $\mathbf{B}_{\textrm{ext}}=\nabla \times \mathbf{A}_{\textrm{ext}}(\mathbf{r})=\mathbf{e}_z B$ in $z$-direction and $\mathbf{A}_{\textrm{ext}}(\mathbf{r})$ is chosen to be in the Landau gauge $\mathbf{A}_{\textrm{ext}}(\mathbf{r})=-\mathbf{e}_x B y$~\cite{Landau}.
The quantized vector potential $\hat{\mathbf{A}}(\bi{r})$ of the electromagnetic field in the Coulomb gauge is~\cite{spohn2004, greiner1996} \begin{equation}\label{AinCoulomb}
\hat{\bi{A}}(\bi{r})=\sum_{\bm{\kappa},\lambda}\sqrt{\frac{\hbar}{\epsilon_0V2\omega(\bm{\kappa})}}\left[ \hat{a}_{\bm{\kappa},\lambda}\bi{S}_{\bm{\kappa},\lambda}(\bi{r})+\hat{a}^{\dagger}_{\bm{\kappa},\lambda}\bi{S}^*_{\bm{\kappa},\lambda}(\bi{r})\right],
\end{equation}
with $\bm{\kappa}=(\kappa_x,\kappa_y,\kappa_z)$ the wave vectors of the photon field, $\omega(\bm{\kappa})=c|\bm{\kappa}|$ the allowed frequencies in the quantization volume $V=L^2L_z$, $\lambda=1,2$ the two transverse polarization directions and $\bi{S}_{\bm{\kappa},\lambda}(\bi{r})$ the vector-valued mode functions, chosen such that the Coulomb gauge is satisfied, $\nabla\cdot\bi{S}_{\bm{\kappa},\lambda}(\bi{r})=0$ ~\cite{spohn2004, greiner1996}. In order for the mode functions $\bi{S}_{\bm{\kappa},\lambda}(\bi{r})$ to satisfy the boundary conditions of the cavity, the wave vectors of the photon field can only take the values $\bm{\kappa}=(\kappa_x,\kappa_y,\kappa_z)=(2\pi n_x/L,2\pi n_y/L,\pi n_z/L_z)$ with $\bi{n}=(n_x,n_y,n_z)\in\mathbb{Z}^3$. The operators $\hat{a}_{\bm{\kappa},\lambda}$ and $\hat{a}^{\dagger}_{\bm{\kappa},\lambda}$ are the annihilation and creation operators of the photon field and obey the bosonic commutation relations $[\hat{a}_{\bm{\kappa},\lambda},\hat{a}^{\dagger}_{\bm{\kappa}^{\prime},\lambda^{\prime}}]=\delta_{\bm{\kappa}\bm{\kappa}^{\prime}}\delta_{\lambda\lambda^{\prime}}$. We note that the annihilation and creation operators can also be defined in terms of the displacement coordinates $q_{\bm{\kappa},\lambda}$ and their conjugate momenta $\partial / \partial q_{\bm{\kappa},\lambda}$ as $\hat{a}_{\bm{\kappa},\lambda}=\frac{1}{\sqrt{2}}\left(q_{\bm{\kappa},\lambda}+\partial/\partial q_{\bm{\kappa},\lambda}\right)\; \textrm{and} \; \hat{a}^{\dagger}_{\bm{\kappa},\lambda}=\frac{1}{\sqrt{2}}\left(q_{\bm{\kappa},\lambda}-\partial/\partial q_{\bm{\kappa},\lambda}\right)$.  

It is clear that in the Pauli-Fierz Hamiltonian both the quantized field $\hat{\bi{A}}(\bi{r})$ defined in Eq.~(\ref{AinCoulomb}) as well as the external vector potential $\bi{A}_{\textrm{ext}}(\bi{r})$ that induces the perpendicular, homogeneous magnetic field break translational symmetry because they are spatially inhomogeneous. This implies that Bloch's theorem cannot be applied for the general Pauli-Fierz Hamiltonian.
However, a way to restore translational symmetry in the full electron-photon configuration space was found after performing the long-wavelength limit (or dipole approximation) for the quantized field $\hat{\bi{A}}(\bi{r})$~\cite{rokaj2019}. In the long-wavelength limit~\cite{rokaj2017, faisal1987}, which has been proven adequate for cavity QED systems~\cite{schafer2019modification,ruggenthaler2017b}, the mode functions $\bi{S}_{\bm{\kappa},\lambda}(\bi{r})$ become spatially independent vectors $\bi{S}_{\bm{\kappa},\lambda}(\bi{r})=\bm{\varepsilon}_{\lambda}(\bm{\kappa})$ and satisfy the condition $\bm{\varepsilon}_{\lambda}(\bm{\kappa})\cdot \bm{\varepsilon}_{\lambda^{\prime}}(\bm{\kappa})=\delta_{\lambda\lambda^{\prime}}$. In addition, we keep only a single-mode of the quantized field and the polarization of the dipolar quantized field is chosen to be parallel to the external vector potential. Under these assumptions the Pauli-Fierz Hamiltonian simplifies to
\begin{eqnarray}\label{velgauge 1mode} 
\hat{H}&=&\sum\limits^{N}_{j=1}\left[\frac{1}{2\me}\left(\mathrm{i}\hbar \mathbf{\nabla}_{j}+e\hat{\mathbf{A}}+e\mathbf{A}_{\textrm{ext}}(\bi{r}_j)\right)^2 +v_{\textrm{ext}}(\mathbf{r}_{j})\right]\nonumber\\
&+&\frac{1}{4\pi\epsilon_0}\sum\limits^{N}_{j< k}\frac{e^2}{|\mathbf{r}_j-\mathbf{r}_k|}+\hbar\omega\left(\hat{a}^{\dagger}\hat{a}+\frac{1}{2}\right)
\end{eqnarray}
and respectively the quantized field in the dipole approximation is~\cite{spohn2004}
\begin{equation}\label{eq2.4b}
\hat{\bi{A}}=\left(\frac{\hbar}{\epsilon_0 V}\right)^{\frac{1}{2}}\frac{\bi{e}_x}{\sqrt{2\omega}}\left( \hat{a}+\hat{a}^{\dagger}\right).
\end{equation}
This single-mode quantized field describes the cavity mode which is strongly coupled to the matter system. However, in Appendix~\ref{Many Modes} we will also take into account the effect of many modes for the integer Hall effect inside a cavity.

Translational symmetry can be restored for the combined electron-photon system in the optical limit (or dipole approximation) $\omega\to 0$~\cite{rokaj2019}.\footnote{We note that the limit $\omega\rightarrow 0$  does not imply that the photonic Hamiltonian $\hat{H}_p$ is identically zero.} To be able to perform consistently this limit one needs to treat exactly and non-perturbatively the $\hat{\bi{A}}^2$ term of the quantized mode. To do so, we isolate the purely photonic part of the Pauli-Fierz Hamiltonian, namely the part that depends only on the photonic annihilation and creation operators $\hat{a}$ and $\hat{a}^{\dagger}$,
\begin{equation}
\hat{H}_p=\hbar\omega\left(\hat{a}^{\dagger}\hat{a}+\frac{1}{2}\right)+\frac{Ne^2}{2\me}\hat{\mathbf{A}}^2.
\end{equation}
The purely photonic part can be brought into the form of the standard harmonic oscillator as follows. The diamagnetic $\hat{\bi{A}}^2$ term renormalizes the photon frequency $\omega$ by introducing the diamagnetic shift $\omega_p=\sqrt{e^2 n_e/\me\epsilon_0}$ which depends on the electron density $n_e$. Then we introduce the scaled coordinate $u=q\sqrt{\widetilde{\omega}/\omega}$ where the dressed photon frequency $\widetilde{\omega}$ is defined by  $\widetilde{\omega}^2=\omega^2+\omega^2_p$.
In terms of the new coordinate $u$ and its conjugate momentum $\partial_u$ the photonic part $\hat{H}_p$ takes the form of a simple harmonic oscillator with frequency $\widetilde{\omega}$,
\begin{equation}
\hat{H}_p=\frac{\hbar \widetilde{\omega}}{2}\left(-\frac{\partial^2}{\partial u^2}+u^2\right).
\end{equation}
The quantized photon field in terms of $u$ is
\begin{equation}
 \hat{\bi{A}}=\sqrt{\frac{\hbar}{\epsilon_0 V\widetilde{\omega}}}u\mathbf{e}_x
\end{equation}
and performing now the optical limit $\omega\rightarrow 0$ we find that the dressed frequency $\widetilde{\omega}$ simply goes to the diamagnetic frequency $\omega_p$. Substituting the expressions for the purely photonic part $\hat{H}_p=\left(\hbar\omega_p/2\right)\left(-\partial^2_u+u^2\right)$ and the vector potential
\begin{equation}
\hat{\mathbf{A}}=\sqrt{\frac{\hbar}{\epsilon_0V\omega_p}} u \mathbf{e}_x,
\end{equation}
back into~(\ref{velgauge 1mode}) we obtain the Pauli-Fierz Hamiltonian in the optical limit
\begin{eqnarray}\label{optical} 
\hat{H}&=&\sum\limits^{N}_{j=1}\left[\frac{1}{2\me}\left(\mathrm{i}\hbar \mathbf{\nabla}_{j}+e\hat{\mathbf{A}}+e\mathbf{A}_{\textrm{ext}}(\bi{r}_j)\right)^2 +v_{\textrm{ext}}(\mathbf{r}_{j})\right]\nonumber\\
&+&\frac{1}{4\pi\epsilon_0}\sum\limits^{N}_{j< k}\frac{e^2}{|\mathbf{r}_j-\mathbf{r}_k|}-\frac{\hbar\omega_p}{2}\frac{\partial^2}{\partial u^2}\;.
\end{eqnarray}
We note that in order to obtain the above expression for the Hamiltonian we used that $\hbar\omega_pu^2/2=Ne^2\hat{\bi{A}}^2/2\me$.\\

Let us now check the translational properties of the Hamiltonian given by Eq.~(\ref{optical}). For simplicity and to avoid specifying a certain lattice geometry, we will consider the case of no external potential, $v_{\textrm{ext}}(\bi{r})=0$. The Hamiltonian is not periodic in the electronic coordinates because $\mathbf{A}_{\textrm{ext}}(\mathbf{r})$ is linear in $y$. However, the Hamiltonian in the optical limit is periodic under a generalized translation in the full electronic plus photonic configuration space,
\begin{equation}\label{symmetry}
(\mathbf{r}_j,u)\longrightarrow\left(\mathbf{r}_j+\mathbf{a},u+Ba_y\sqrt{\epsilon_0 V\omega_p/\hbar}\right),
\end{equation}
where $\bi{a}=(a_x,a_y,a_z) \in \mathbb{R}^3$ arbitrary. This is true because the total vector potential $\hat{\bi{A}}_{\textrm{tot}}=\hat{\bi{A}}+\bi{A}_{\textrm{ext}}(\bi{r}_j)$ is invariant under the above generalized translation and obviously the kinetic terms of the quantized mode and of the electrons are also invariant, as well as the Coulomb interaction. The fact that the Hamiltonian in Eq.~(\ref{optical}) is invariant under the continuous translations of Eq.~(\ref{symmetry}) implies that $\hat{H}$ will also be invariant under Bravais lattice translations in the case of a periodic potential.

\section{Effective Hamiltonian \& QED-Bloch Expansion}\label{QED Bloch Expansion}

Having restored translational symmetry, our goal now is to go one step further and construct a Bloch-type ansatz in the polaritonic (electronic plus photonic) configuration space and to derive a Bloch-type central equation for the description of solids in a classical, homogeneous magnetic field coupled to a quantized electromagnetic field. 

To make the problem tractable, instead of treating the unfeasible many-body interacting Hamiltonian of Eq.~(\ref{optical}), we will employ the independent electron approximation which is commonly used in condensed matter physics. We note that this independent-electron approach is consistent with Bloch theory, which is not a theory of a single electron in a periodic potential but rather of many non-interacting electrons. 

To incorporate the fact that the charged particles couple collectively to the photon field, we will use an effective electron density $n_e$ and we will scale the strength of the quantized field $\hat{\bi{A}}$ by the square root of the number of charges $\sqrt{N}$,
\begin{equation}\label{effective vector potential}
\hat{\bi{A}} \longrightarrow \hat{\bi{\mathcal{A}}}=\sqrt{N}\hat{\bi{A}}=\sqrt{\frac{\hbar \omega_p \me}{e^2}}u\bi{e}_x.
\end{equation}
Keeping an effective electron density and scaling the field by $\sqrt{N}$ allows us to capture the back-reaction of matter to the photon field and to describe the emergence of novel quasiparticle excitations known as Landau polaritons~\cite{rokaj2019}. We would also like to mention that the scaling of light-matter interaction by $\sqrt{N}$ is the standard argument for the description of collective coupling in the few-level models of quantum optics~\cite{JoelCollective,dicke1954, garraway2011, kockum2019ultrastrong}. For the inclusion of any further effects, like exchange and correlation effects, one would need the addition of effective fields as introduced in quantum-electrodynamical density functional theory (QEDFT)~\cite{ruggenthaler2014, ruggenthaler2015, TokatlyPRL, ChristianFunctional}.

Upon these assumptions, we obtain the following single-particle effective Hamiltonian, 
\begin{equation}\label{effective} 
\Heff=\frac{1}{2\me}\left(\mathrm{i}\hbar \mathbf{\nabla}+e\hat{\bi{\mathcal{A}}}+e\mathbf{A}_{\textrm{ext}}(\bi{r})\right)^2-\frac{\hbar\omega_p}{2}\frac{\partial^2}{\partial u^2}+v_{\textrm{ext}}(\mathbf{r}).
\end{equation}
which was already proposed by the authors in Ref.~\cite{rokaj2019} and was applied successfully to the description of Landau polariton systems. Before we continue, we would like to specify the geometries in which we are interested in this article.

\subsection{Setting the Geometry}

Our aim here is to treat all possible 2D geometries of periodic structures. The external potential in a solid is assumed periodic $v_{\textrm{ext}}(\mathbf{r})=v_{\textrm{ext}}(\mathbf{r}+\mathbf{R}_{\mathbf{n}})$ where $\mathbf{R}_{\mathbf{n}}$ is a Bravais lattice vector with $\mathbf{n}=(n,m)\in \mathbb{Z}^2$. The Bravais lattice vectors in general are $\mathbf{R}_{\mathbf{n}}=n\mathbf{a}_1+m\mathbf{a}$, where $\mathbf{a}_1$ and $\bi{a}_2$ are the primitive vectors which lie in different directions and span the 2D lattice. Without loss of generality we can choose the vector $\mathbf{a}_1$ to be in the $x$-direction $\mathbf{a}_1=a_1\mathbf{e}_x$. The second primitive vector in this case is $\mathbf{a}_2=a_2\cos\theta\mathbf{e}_x+a_2\sin\theta\mathbf{e}_y$, where $\theta$ is the angle between the vectors $\mathbf{a}_1$ and $\mathbf{a}_2$. Thus, the Bravais lattice vectors are 
\begin{equation}\label{3DBravaisvectors}
\mathbf{R}_{\mathbf{n}}=\left(na_1+ma_2\cos\theta\right)\mathbf{e}_x+ma_2\sin\theta\mathbf{e}_y.
\end{equation}
Then, the reciprocal lattice vectors are $\mathbf{G}_{\mathbf{n}}=n\mathbf{b}_1+m\mathbf{b}_2$ with $\mathbf{n}=(n,m)\in \mathbb{Z}^2$. The defining relation for the vectors $\mathbf{b}_{1}$ and $\bi{b}_2$ is $\mathbf{b}_{i}\cdot \mathbf{a}_j=2\pi \delta_{ij} \;\; \textrm{with}\;\; i,j=1,2$~\cite{Mermin, Callaway}. With the given choice of primitive vectors the reciprocal primitive vectors are
\begin{equation}
\mathbf{b}_1=\frac{2\pi}{a_1}\mathbf{e}_x-\frac{2\pi\cos\theta}{a_1\sin\theta}\mathbf{e}_y\;\;\textrm{and}\;\; \mathbf{b}_2=\frac{2\pi}{a_2\sin\theta}\mathbf{e}_y.
\end{equation}
Thus, the reciprocal lattice vectors are
\begin{equation}
\mathbf{G}_{\mathbf{n}}=\frac{2\pi n}{a_1}\mathbf{e}_x +\left(\frac{2\pi m}{a_2\sin\theta}-\frac{2\pi n\cos\theta}{a_1\sin\theta}\right) \mathbf{e}_y,
\end{equation}
which for convenience we will write as
\begin{eqnarray}\label{3Dreciprocallattice}
&&\mathbf{G}_{\mathbf{n}}=\left(G^x_{n}, G_{\mathbf{n}}\right)\;\; \textrm{with}\;\;G_{\mathbf{n}}=-\frac{G^x_{n}\cos\theta}{\sin\theta}+\frac{G^y_{m}}{\sin\theta}\nonumber\\
 &&\textrm{and}\;\;  G^x_{n}=\frac{2\pi n}{a_1 },\;\;\;G^y_{m}=\frac{2\pi m}{a_2}.
\end{eqnarray}
With these choices we have defined our geometrical setting and we have made clear how all possible 2D Bravais lattices can be described. 

\subsection{Polaritonic Coordinates}

To continue, we introduce the cyclotron frequency $\omega_c=eB/\me$ and we expand the covariant kinetic term of the effective Hamiltonian into
\begin{eqnarray}\label{Approptical}
\Heff=&-&\frac{\hbar^2}{2\me}\nabla^2+\mathrm{i}\hbar\mathbf{e}_x\left(u\sqrt{\hbar\omega_p/\me}-y\omega_c\right)\cdot\nabla\\
&+&v_{\textrm{ext}}(\mathbf{r})+\frac{\me}{2}\left(u\sqrt{\hbar\omega_p/\me}-y\omega_c\right)^2-\frac{\hbar\omega_p}{2}\frac{\partial^2}{\partial u^2}.\nonumber
\end{eqnarray}
This effective Hamiltonian is invariant under a translation that acts on both the electronic and photonic coordinates for any 2D Bravais lattice vector $\bi{R}_{\bi{n}}$,
\begin{equation}\label{1psymmetry}
(\mathbf{r},u)\longrightarrow \left(\mathbf{r}+\mathbf{R}_{\mathbf{n}},u+ma_2\sin\theta \omega_c \sqrt{\me/\hbar\omega_p}\right).
\end{equation}
To describe properly this symmetry in the polaritonic space, we will switch to a new set of coordinates. For that purpose, we introduce the relative distance and center-of-mass coordinates between rescaled versions of $u$ and $y$,
\begin{equation}\label{w and v coordinates}
w=\frac{m_pu\sqrt{\hbar\omega_p/\me}+m_c\omega_cy}{\sqrt{2}M},\; v=\frac{u\sqrt{\hbar\omega_p/\me}-\omega_cy}{\sqrt{2}},
\end{equation}
and the Hamiltonian $\Heff$ simplifies to
\begin{equation}
\begin{aligned}
\Heff&=-\frac{\hbar^2}{2\me}\frac{\partial^2}{\partial x^2} -\frac{\hbar^2}{2M}\frac{\partial^2}{\partial w^2}+v_{\textrm{ext}}(\bi{r})\\ &+\textrm{i}\hbar\sqrt{2}v\frac{\partial}{\partial x} +\me v^2-\frac{\hbar^2}{2\mu}\frac{\partial^2}{\partial v^2},
\end{aligned}
\end{equation}
with the mass parameters $M, \mu, m_p$ and $m_c$ being
\begin{equation}\label{mass polaritonic parameters}
\begin{aligned}
m_{p}&=\frac{\me}{\omega^2_p}, \;\;\; m_c=\frac{\me}{\omega^2_c},\\ M&=\frac{m_p+m_c}{2}\;\;\; \textrm{and}\;\;\; \mu=\frac{m_pm_c}{M}.
\end{aligned}
\end{equation}
Furthermore, by performing a square completion, the effective Hamiltonian can be written in the compact form
\begin{equation}\label{Heff compact}
\begin{aligned}
\Heff=&-\frac{\hbar^2}{2M}\frac{\partial^2}{\partial w^2}-\frac{\hbar^2}{2\mu}\frac{\partial^2}{\partial v^2 }\\
&+ \frac{\mu\Omega^2}{2}\left(v+\frac{\textrm{i}\hbar}{\sqrt{2}\me}\frac{\partial}{\partial x}\right)^2+ v_{\textrm{ext}}(\bi{r}),
\end{aligned}
\end{equation}
where the dressed frequency $\Omega$ is defined by
\begin{equation}\label{upper polariton frequency}
\Omega^2=\frac{2\me}{\mu}=\omega^2_p+\omega^2_c.
\end{equation}
The original electronic vector $\bi{r}=(x,y)$ in the new polaritonic coordinate system is
\begin{equation}
\bi{r}=(x,y)=\left(x,\frac{w}{\sqrt{2}\omega_c}-\frac{m_p v}{\sqrt{2}M\omega_c}\right).
\end{equation}
It is important to note that the coordinates $v$ and $w$ are independent, because their respective position and momentum operators commute. Moreover, we would like to emphasize that since the external potential is periodic, it can be written in terms of a Fourier series, which in terms of the polaritonic coordinates is
\begin{equation}\label{potential Fourier}
v_{\textrm{ext}}(\bi{r})=\sum_{\bi{n}}V_{\bi{n}} \rme^{\textrm{i}\bi{G}_{\bi{n}}\cdot \bi{r}}=\sum_{\bi{n}}V_{\bi{n}}\rme^{\textrm{i}\bi{G}^w_{\bi{n}}\cdot \bi{r}_w}\rme^{-\textrm{i}G^v_{\mathbf{n}}v},
\end{equation}
where
\begin{eqnarray}\label{pol reciprocal lattice vectors }
&&G^v_{\mathbf{n}}=\frac{m_pG_{\mathbf{n}}}{\sqrt{2}M\omega_c}, \quad \bi{r}_w=(x,w)\;\; \textrm{and}\\
&& \bi{G}^w_{\bi{n}}=(G^x_{n},G^w_{\mathbf{n}})=(G^x_{n},G_{\mathbf{n}}/\sqrt{2}\omega_c).\nonumber
\end{eqnarray}

\subsection{QED-Bloch Expansion}

The Hamiltonian $\Heff$ of Eq.~(\ref{Heff compact}) is invariant under translations in the polaritonic configuration space 
\begin{equation}\label{polariton Bravais}
(x,w)\longrightarrow (x+na_1+ma_2\cos\theta,w+m\sqrt{2}\omega_ca_2\sin\theta).
\end{equation}
This implies that we can use Bloch's theorem in the $(x,w)$ plane. Consequently, the eigenfunctions of $\Heff$ can be written with the ansatz 
\begin{equation}\label{BlochAnsatz}
\Psi_{\mathbf{k}}(\mathbf{r}_{w},v)=\rme^{\mathrm{i}\mathbf{k}\cdot\mathbf{r}_w}U^{\mathbf{k}}(\mathbf{r}_w,v)
\end{equation}
where $\mathbf{r}_w=(x,w)$. Here the function $U^{\mathbf{k}}(\mathbf{r}_w,v)$ is periodic under the translations in the polaritonic space defined in Eq.~(\ref{polariton Bravais}). The crystal momentum $\mathbf{k}=(k_x,k_w)$ corresponds to $\mathbf{r}_w$ and $k_w$ is a polaritonic quantum number. Note that the polaritonic unit cell in the $w$-direction scales linearly with the strength of the magnetic field. The same feature appears also for the usual magnetic unit cell, but in this case only field strengths which generate a rational magnetic flux through a unit cell are allowed~\cite{Kohmoto}. Contrary to that, the polaritonic unit cell puts no restrictions on the strength of the magnetic field.

Since the function $U^{\mathbf{k}}(\mathbf{r}_w,v)$ is periodic in $\mathbf{r}_w$ we expand it in a Fourier series in $\mathbf{r}_w$, while for the $v$-dependent part we consider a generic function,
\begin{equation}
\Psi_{\mathbf{k}}(\mathbf{r}_{w},v)=\rme^{\mathrm{i}\mathbf{k}\cdot\mathbf{r}_w}\sum_{\mathbf{n}}U^{\mathbf{k}}_{\mathbf{n}}\rme^{\mathrm{i}\mathbf{G}^w_{\mathbf{n}}\cdot\mathbf{r}_w}\phi^{\bi{k}}_{\bi{n}}(v),
\end{equation}
where $\mathbf{G}^w_{\mathbf{n}}=(G^x_n,G^w_{\mathbf{n}})$ are the reciprocal lattice vectors in the $(x,w)$-space. We substitute the above ansatz wavefunction into the Schrödinger equation with $\Heff$ and we get
\begin{widetext}
\begin{equation}
\sum_{\bi{n}}U^{\bi{k}}_{\bi{n}}\rme^{\textrm{i}(\bi{k}+\bi{G}^w_{\bi{n}})\cdot\bi{r}_w} \Bigg[\frac{\hbar^2(k_w+G^w_{\bi{n}})^2}{2M}+v_{\textrm{ext}}(\bi{r})-\frac{\hbar^2}{2\mu}\frac{\partial^2}{\partial v^2}+\frac{\mu \Omega}{2}\left(v-\frac{\hbar(k_x+G^x_n)}{\sqrt{2}\me}\right)^2-E_{\bi{k}}\Bigg]\phi^{\bi{k}}_{\bi{n}}(v)=0.
\end{equation}
\end{widetext}
The Hamiltonian includes a harmonic-oscillator part $\hat{H}_v$ which is shifted by the momentum in the $x$-direction
\begin{equation}\label{HO-shift}
A^{k_x}_n=\frac{\hbar(k_x+G^x_n)}{\sqrt{2}\me},
\end{equation}
written out as
\begin{equation}
\hat{H}_v=-\frac{\hbar^2}{2\mu}\frac{\partial^2}{\partial v^2}+\frac{\mu \Omega^2}{2}\left(v-A^{k_x}_n\right)^2.
\end{equation}
The eigenfunctions of this operator are the shifted Hermite functions $\phi_j(v-A^{k_x}_n)$ with eigenenergies 
\begin{equation}\label{Landau polariton energies}
\begin{aligned}
\mathcal{E}_j&=\hbar\Omega\left(j+\frac{1}{2}\right)=\hbar\sqrt{\omega^2_p+\omega^2_c}\left(j+\frac{1}{2}\right)\\
&=\hbar\omega_c\sqrt{1+\eta^2}\left(j+\frac{1}{2}\right)
\end{aligned}
\end{equation}
which are degenerate with respect to the momentum in $x$-direction. The eigenstates $\phi_{j} \left(v-A^{k_x}_n\right)$, as it was shown in Ref.~\cite{rokaj2019}, correspond to Landau polaritons~\cite{Keller2020} and have many structural similarities to the well-known Landau levels~\cite{Landau}. Further, the Landau polariton energy levels can be understood as Landau levels modified by the coupling to the cavity
\begin{equation}\label{fractal coupling}
\eta=\frac{\omega_p}{\omega_c},
\end{equation}
which in our setting is defined as the ratio between the two fundamental scales in the light-matter coupled system, namely the diamagnetic frequency $\omega_p$ and the cyclotron frequency $\omega_c$. For $\eta=0$ the Landau polariton energy levels reduce to the standard Landau levels~\cite{Landau}.

 We will now make use of the Landau polariton eigenfunctions by expanding $\phi^{\bi{k}}_{\bi{n}}(v)$ in terms of this basis. Then, the polaritonic Bloch ansatz takes the form 
\begin{equation}\label{BlochAnsatzHermite}
\Psi_{\mathbf{k}}(\mathbf{r}_{w},v)=\rme^{\mathrm{i}\mathbf{k}\cdot\mathbf{r}_w}\sum_{\mathbf{n},j}U^{\mathbf{k}}_{\mathbf{n},j}\rme^{\mathrm{i}\mathbf{G}^w_{\mathbf{n}}\cdot\mathbf{r}_w}\phi_j(v-A^{k_x}_n).
\end{equation} 
With respect to our previous work~\cite{rokaj2019} here we use shifted Hermite functions $\phi_j(v-A^{k_x}_n)$ for the construction of our ansatz, instead of unshifted ones. The shifted basis helps to incorporate the degeneracy of the Landau polaritons with respect to $k_x$.

Substituting the polaritonic Bloch ansatz again into our Schr\"{o}dinger equation and making use of the Fourier expansion of the external potential given in Eq.~(\ref{potential Fourier}) we have
\begin{eqnarray}
&&\sum_{\bi{n},j}U^{\bi{k}}_{\bi{n},j}\rme^{\textrm{i}\bi{G}^w_{\bi{n}}\cdot\bi{r}_w} \phi_j\left(v-A^{k_x}_n\right) \Bigg[\frac{\hbar^2(k_w+G^w_{\bi{n}})^2}{2M}+\mathcal{E}_j-E_{\bi{k}}\Bigg]\nonumber\\
&&+\sum_{\bi{n},\bi{n}^{\prime},j}V_{\bi{n}^{\prime}} U^{\mathbf{k}}_{\mathbf{n},j} \rme^{\textrm{i}\bi{G}^w_{\bi{n}+\bi{n}^{\prime}}\cdot \bi{r}_w}\rme^{-\textrm{i}G^v_{\bi{n}^{\prime}}v}\phi_j(v-A^{k_x}_n)=0.
\end{eqnarray}
To eliminate the plane waves depending on $\bi{r}_w$ we multiply the above expression by $\rme^{-\textrm{i}\bi{G}^w_{\bi{q}}\cdot \bi{r}_w}$ and integrate over $\bi{r}_w$,
\begin{eqnarray}
&&\sum_{j}U^{\bi{k}}_{\bi{n},j} \phi_j\left(v-A^{k_x}_n\right) \Bigg[\frac{\hbar^2(k_w+G^w_{\bi{n}})^2}{2M}+\mathcal{E}_j-E_{\bi{k}}\Bigg]\nonumber\\
&&+\sum_{\bi{n}^{\prime},j}V_{\bi{n}-\bi{n}^{\prime}} U^{\mathbf{k}}_{\mathbf{n}^{\prime},j} \rme^{-\textrm{i}G^v_{\bi{n}-\bi{n}^{\prime}}v}\phi_j(v-A^{k_x}_{n^{\prime}})=0.
\end{eqnarray}
We note that after the integration we exchanged the index $\bi{q}$ with $\bi{n}$ again. Next, we apply the bra\footnote{We note that the standard bra and ket notation does not include the coordinate. Here, we kept the coordinate as it will be convenient to perform shift transformations on these states to obtain the matrix representation of the displacement operators. See Appendix~\ref{Displacement Algebra} for this.} $\langle \phi_i(v-A^{k_x}_n)|$ from the left,
\begin{eqnarray}\label{beforeMatrixeq}
&&0=U^{\bi{k}}_{\bi{n},i} \Bigg[\frac{\hbar^2(k_w+G^w_{\bi{n}})^2}{2M}+\mathcal{E}_i-E_{\bi{k}}\Bigg]+\\
&&\sum_{\bi{n}^{\prime},j}V_{\bi{n}-\bi{n}^{\prime}} U^{\mathbf{k}}_{\mathbf{n}^{\prime},j}\; \langle \phi_i(v-A^{k_x}_n)|\rme^{-\textrm{i}G^v_{\bi{n}-\bi{n}^{\prime}}v}|\phi_j(v-A^{k_x}_{n^{\prime}})\rangle.\nonumber
\end{eqnarray}
The only thing left to be computed in order to obtain our QED-Bloch central equation are the matrix elements
\begin{equation}\label{matrix elements}
\langle \phi_i(v-A^{k_x}_n)|\rme^{-\textrm{i}G^v_{\bi{n}-\bi{n}^{\prime}}v}|\phi_j(v-A^{k_x}_{n^{\prime}})\rangle.
\end{equation}
This can be performed with the use of displacement operators and their corresponding algebra~\cite{Glauber}. We present this derivation in Appendix~\ref{Displacement Algebra} and we find 
\begin{eqnarray}\label{Matrixeq}
&&\langle \phi_i(v-A^{k_x}_n)|\rme^{-\textrm{i}G^v_{\bi{n}-\bi{n}^{\prime}}v}|\phi_j(v-A^{k_x}_{n^{\prime}})\rangle\\
&&= \rme^{-\textrm{i}G^v_{\bi{n}-\bi{n}^{\prime}}A^{k_x}_{(n+n^{\prime})/2}}\langle \phi_i|\hat{D}(\alpha_{\bi{n}-\bi{n}^{\prime}})|\phi_j\rangle\nonumber
\end{eqnarray}
where $\hat{D}(\alpha_{\bi{n}-\bi{n}^{\prime}})$ is a displacement operator with the shift given by
\begin{equation}\label{alphamatrix}
\alpha_{\bi{n}-\bi{n}^{\prime}}=-\sqrt{\frac{\mu\Omega}{2\hbar}}A^0_{n-n^{\prime}}-\textrm{i}\sqrt{\frac{\hbar}{2\mu\Omega}}G^v_{\bi{n}-\bi{n}^{\prime}}.
\end{equation}
Further, the matrix representation of this displacement operator in the basis $\{\phi_i\}$ is~\cite{Glauber}
\begin{eqnarray}\label{displacementeq}
\langle \phi_i|\hat{D}(\alpha_{\bi{n}-\bi{n}^{\prime}})|\phi_j\rangle=\sqrt{\frac{j!}{i!}}\alpha^{i-j}_{\bi{n}-\bi{n}^{\prime}}\rme^{-\frac{|\alpha_{\bi{n}-\bi{n}^{\prime}}|^2}{2}}L^{(i-j)}_j(|\alpha_{\bi{n}-\bi{n}^{\prime}}|^2)\nonumber\\
\end{eqnarray}
where $i\geq j$ and $L^{(i-j)}_j(|\alpha_{\bi{n}-\bi{n}^{\prime}}|^2)$ are the associated Laguerre polynomials. We note that for $j>i$ one needs to take 
\begin{equation}
\langle \phi_i|\hat{D}(\alpha_{\bi{n}-\bi{n}^{\prime}})|\phi_j\rangle=(-1)^{j-i}\langle \phi_j|\hat{D}(\alpha_{\bi{n}-\bi{n}^{\prime}})|\phi_i\rangle^{*}
\end{equation}
because $\hat{D}^{\dagger}(\alpha)=\hat{D}(-\alpha)$. Finally, substituting Eqs.~\eqref{Matrixeq} and~\eqref{displacementeq} for the matrix representation of the displacement operator into Eq.~\eqref{beforeMatrixeq}, we obtain the QED-Bloch central equation
\begin{widetext}
\begin{equation}\label{QED-Bloch Central}
U^{\bi{k}}_{\bi{n},i} \Bigg[\frac{\hbar^2(k_w+G^w_{\bi{n}})^2}{2M}+\mathcal{E}_i-E_{\bi{k}}\Bigg]+\sum_{\bi{n}^{\prime},j}V_{\bi{n}-\bi{n}^{\prime}} U^{\mathbf{k}}_{\mathbf{n}^{\prime},j}\; \rme^{-\textrm{i}G^v_{\bi{n}-\bi{n}^{\prime}}A^{k_x}_{(n+n^{\prime})/2}}\langle \phi_i|\hat{D}(\alpha_{\bi{n}-\bi{n}^{\prime}})|\phi_j\rangle=0.
\end{equation}
\end{widetext}
The equation above is the main result of QED-Bloch theory. The QED-Bloch central equation provides a unified framework for the description of periodic materials in the presence of homogeneous magnetic fields coupled to a quantized electromagnetic field. It is important to mention that the QED-Bloch central equation is also applicable in the case where there is no quantized field, i.e., where the frequency $\omega_p$ is equal to zero. This implies that QED-Bloch theory and the central equation we derived can be used also for the description of periodic materials solely under the influence of the homogeneous magnetic field. The semi-classical limit of our central equation, where the quantized field is equal to zero, is performed in detail in Appendix~\ref{No quantized field}. In addition, we will also discuss this limit in the context of the Harper equation~\cite{Harper_1955} and the Hofstadter butterfly~\cite{Hofstadter} in Section~\ref{Harper Duality}.

\section{Polaritonic Hofstadter Butterfly}\label{Polaritonic Butterfly}

As a first application of our QED-Bloch theory we consider the case where we have a 2D periodic system perpendicular to a homogeneous external magnetic field and coupled to the quantized mode originating from a cavity as depicted in Fig.~\ref{2D Solid Cavity}. It is important to mention that in such a cavity setting the diamagnetic frequency $\omega_p$ can be defined in terms of the 2D electron density $n_{\textrm{2D}}=N/S$, where $S$ is the area of the 2D material, and the fundamental cavity frequency $\omega_{\textrm{cav}}=\pi c/L_z$ as~\cite{rokaj2019, rokaj2020, EckhardtChain}
\begin{equation}\label{plasma frequency}
    \omega_p=\sqrt{\frac{e^2 n_e}{\me\epsilon_0}}=\sqrt{\frac{e^2n_{\textrm{2D}}\omega_{\textrm{cav}}}{\me\epsilon_0\pi c}}.
\end{equation}

\subsection{Polaritonic Butterfly in Square Lattice \& Polaritonic Harper Equation}

\begin{figure*}
     \centering
     \begin{subfigure}[b]{0.48\textwidth}
         \centering
        \includegraphics[width=\linewidth]{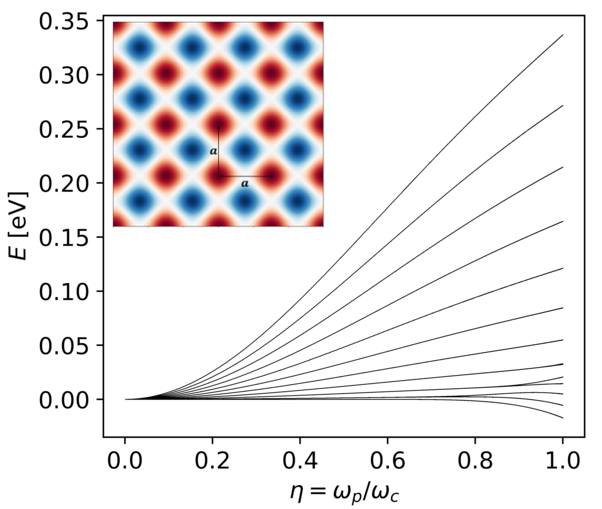}
 \caption{\label{Fractal_pol_0,1}Energy spectrum as a function of the light-matter coupling $\eta=\omega_p/\omega_c$ for magnetic flux ratio $\Phi/\Phi_0=0.1$.}
     \end{subfigure}
     \hfill
     \begin{subfigure}[b]{0.48\textwidth}
         \centering
         \includegraphics[width=\linewidth]{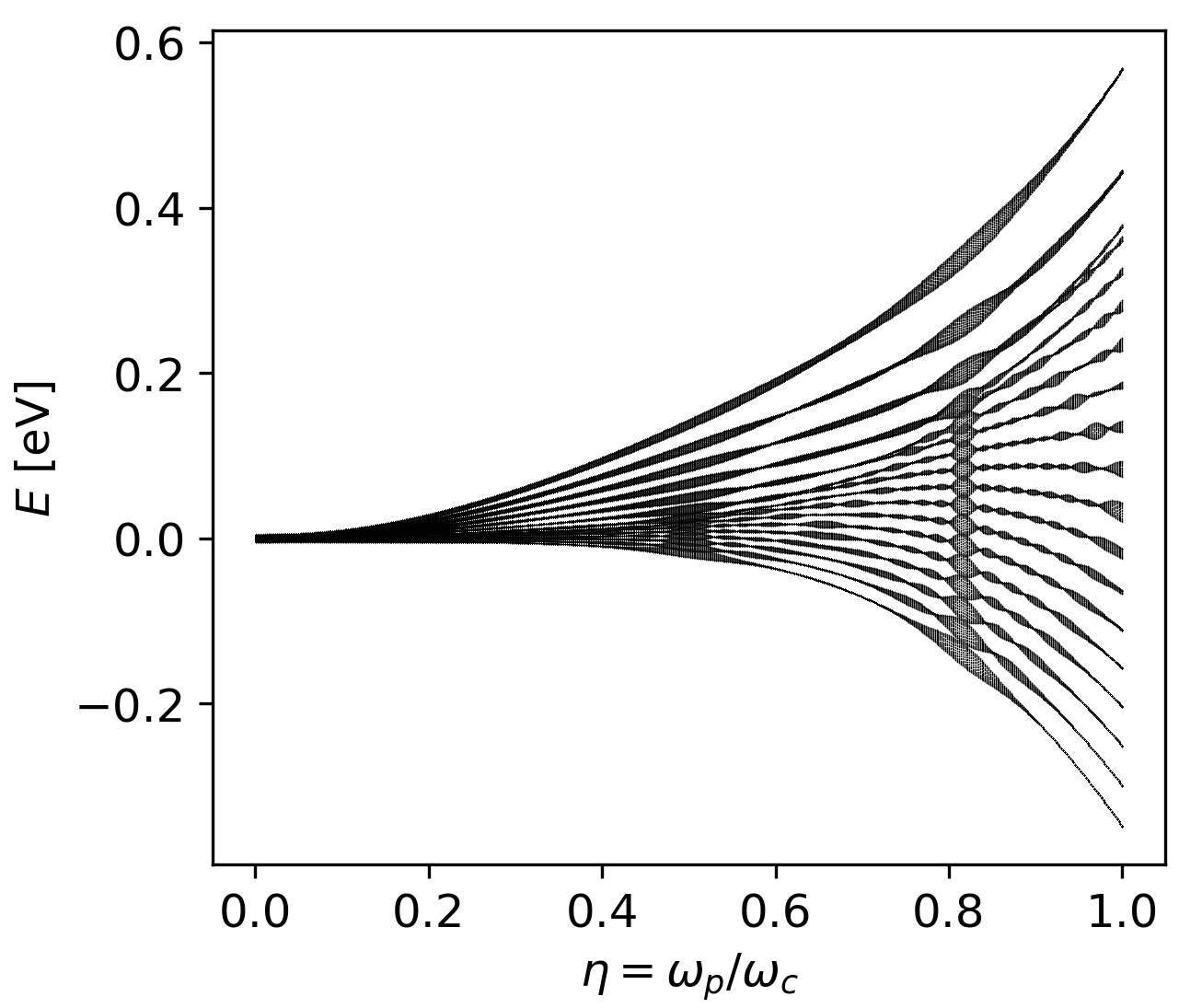}
 \caption{\label{Fractal_pol_0,2}Energy spectrum as a function of the light-matter coupling $\eta=\omega_p/\omega_c$ for magnetic flux ratio $\Phi/\Phi_0=0.2$.}
     \end{subfigure}
     \hfill
      \begin{subfigure}[b]{0.48\textwidth}
         \centering
     \hfill
     \includegraphics[width=\linewidth]{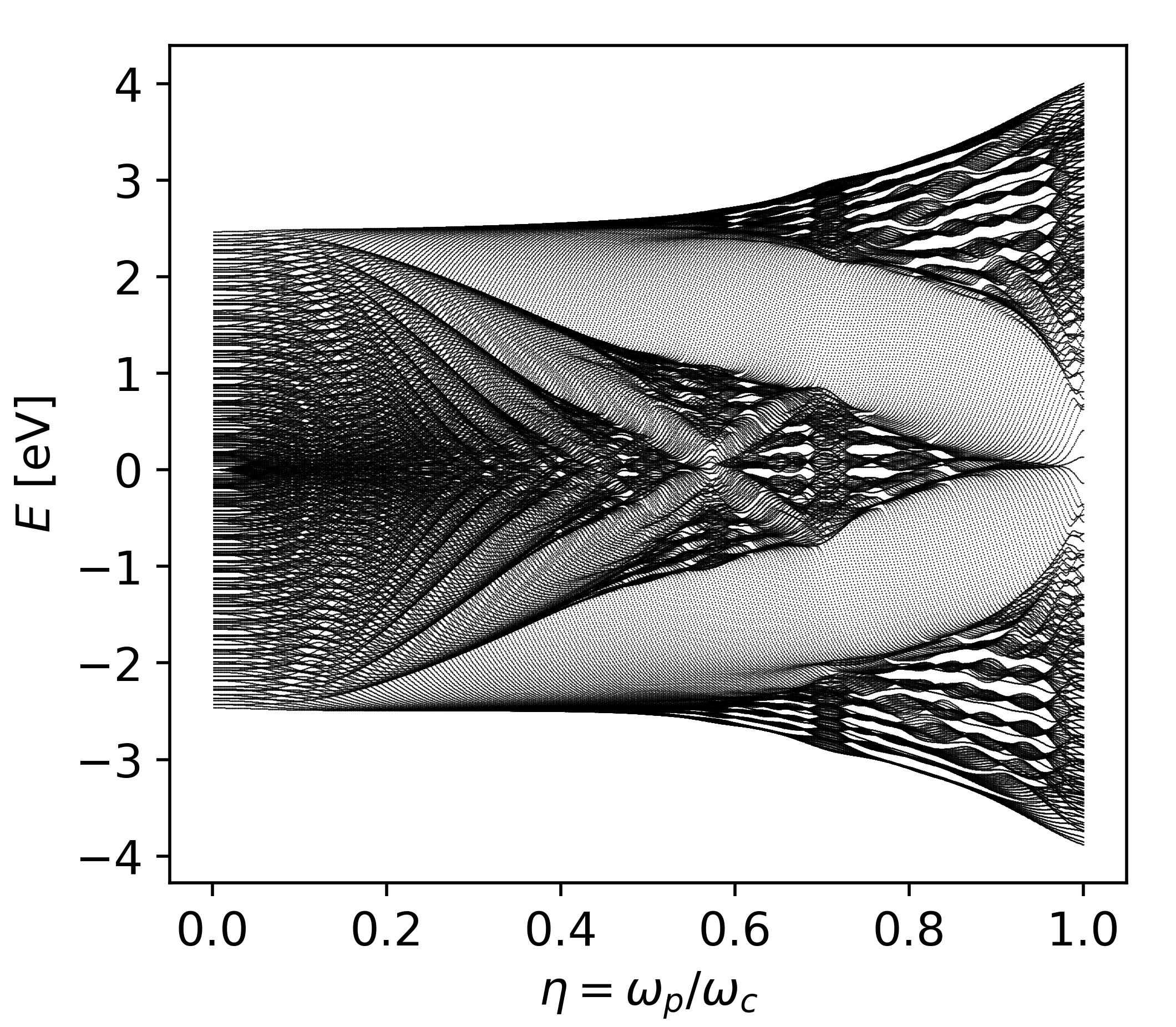}
	\caption{\label{Fractal_pol_1}Energy spectrum as a function of the light-matter coupling $\eta=\omega_p/\omega_c$ for magnetic flux ratio $\Phi/\Phi_0=1$.}
	\end{subfigure}
	\hfill
     \begin{subfigure}[b]{0.48\textwidth}
         \centering
         \hfill
         \includegraphics[width= \linewidth]{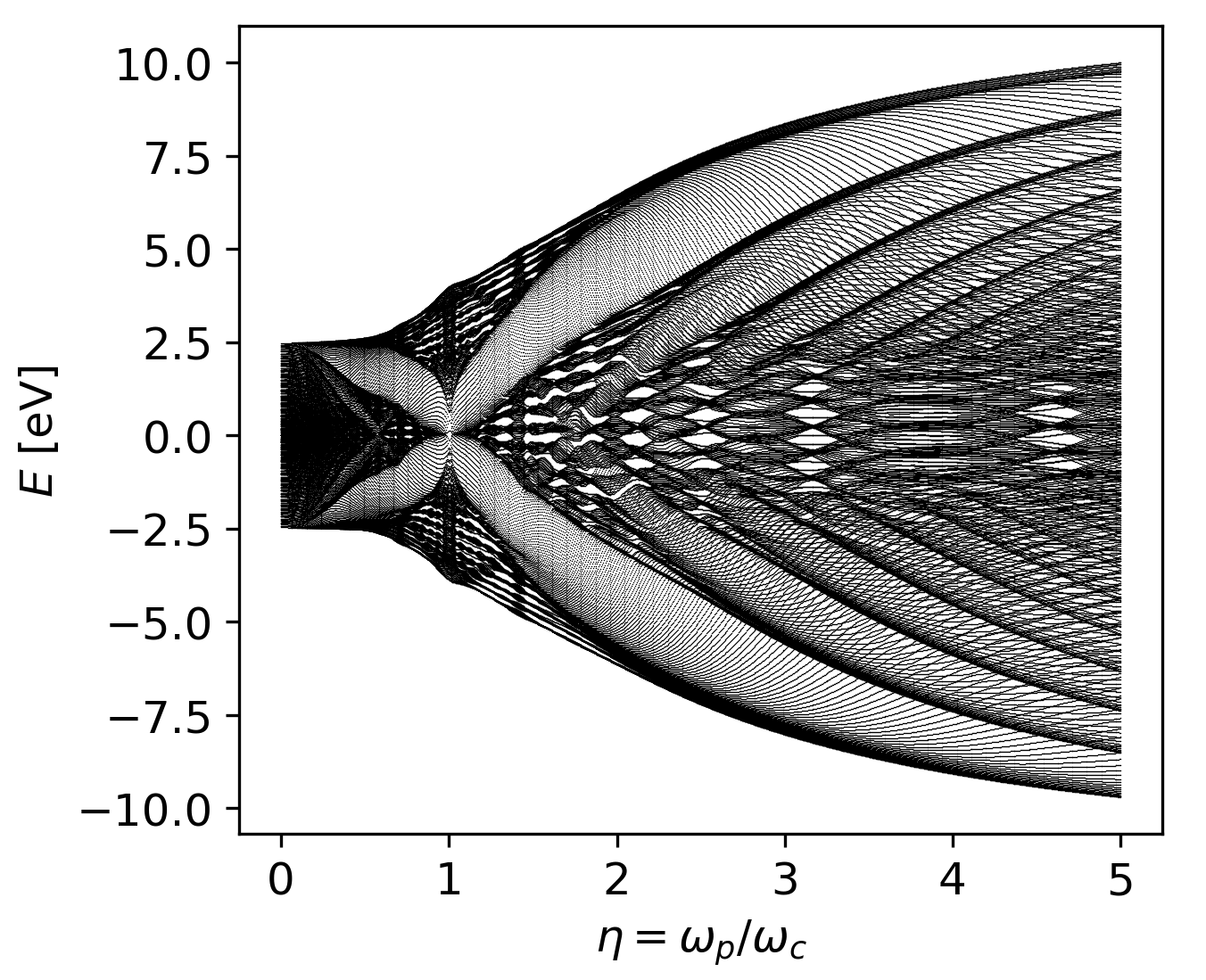}
    \caption{ \label{Fractal_pol_g5}Energy spectrum as a function of the light-matter coupling $\eta=\omega_p/\omega_c$ for magnetic flux ratio $\Phi/\Phi_0=1$.}
     \end{subfigure}
    \caption{Energy spectra as a function of the light-matter coupling $\eta$ for the square cosine potential and for different values of the relative magnetic flux $\Phi/\Phi_0$. The inset in Fig.~\ref{Fractal_pol_0,1} displays the spatial profile of the applied square cosine lattice potential.}
        \label{fig:three graphs}
\end{figure*}

In this subsection, as a first example, we will look into a 2D square cosine potential. For a square lattice potential the angle between the lattice vectors is $\theta=\pi/2$ and the two lattice constants are equal $a_1=a_2=a$. We emphasize that in order to achieve substantial fractions of the magnetic flux quantum $\Phi_0=h/e$ in the magnetic flux $\Phi=Ba^2$, we have to choose the lattice constant of the potential to be 50 times larger than the typical lattice constants in standard materials, which are of the order of a few Ångström. Thus, for our lattice potential the lattice constant is $a=50\times 3 \textrm{\AA}=15\textrm{nm}$. Such enlarged lattice periodicities can be achieved with Moir\'{e} materials~\cite{Moiremarvels, RubioMoire}, and it is within such setups that the experimental demonstration of the Hofstadter butterfly has been achieved~\cite{DeanButterfly, WangButterfly, ForsytheButterfly}.

For the square cosine potential the only non-zero Fourier components are $V_{\pm1,0}=V_{0,\pm1}=V_0$, where $V_0$ determines the strength of the potential that we choose to be $V_0=3\textrm{eV}$. Consequently, our 2D square lattice potential takes the form depicted as an inset in Fig.~\ref{Fractal_pol_0,1}.
For the reciprocal lattice vectors of the square cosine potential it holds $G^v_{\bi{n}}=m_pG^y_m/\sqrt{2}M\omega_c$ and $G^w_{\bi{n}}=G^y_m/\sqrt{2}\omega_c$, see for this Eqs.~(\ref{3Dreciprocallattice}) and (\ref{potential Fourier}). To further simplify our considerations, we also consider the case where only the lowest Landau polariton $\phi_0(v-A^{k_x}_n)$ is occupied. As we will see later, restricting ourselves to the lowest level will help us to connect to the well-known Harper equation~\cite{Harper_1955} and the Hofstadter butterfly~\cite{Hofstadter}.
Under these assumptions, the QED-Bloch central equation takes the simplified form
\begin{widetext}
\begin{equation}
\begin{aligned}
&U^{k_x,k_w}_{n,m} \Bigg[\frac{\hbar^2\left(k_w+\frac{G^y_{m}}{\sqrt{2}\omega_c}\right)^2}{2M}+\mathcal{E}_0-E_{k_x,k_w}\Bigg]+V_0U^{k_x,k_w}_{n-1,m}\rme^{-\frac{|\alpha_{1,0}|^2}{2}}+V_0U^{k_x,k_w}_{n+1,m}\rme^{-\frac{|\alpha_{-1,0}|^2}{2}} \\
&+V_0U^{k_x,k_w}_{n,m-1}\exp\left(\frac{-\textrm{i}m_pG^y_1A^{k_x}_n}{\sqrt{2}M\omega_c}\right)\rme^{-\frac{|\alpha_{0,1}|^2}{2}}+V_0U^{k_x,k_w}_{n,m+1}\exp\left(\frac{-\textrm{i}m_pG^y_{-1}A^{k_x}_n}{\sqrt{2}M\omega_c}\right)\rme^{-\frac{|\alpha_{0,-1}|^2}{2}}=0.
\end{aligned}
\end{equation}
\end{widetext}
For $\theta=\pi/2$ and $a_1=a_2=a$ the $\alpha$-matrix defined in Eq.~(\ref{alphamatrix}) is
\begin{equation}
\alpha_{\bi{n}}=-\frac{2\pi}{a}\sqrt{\frac{\hbar}{2\me\Omega}}\left(n+\textrm{i}\frac{\omega_c}{\Omega}m\right).
\end{equation}
Using the above expression we find for the four components of the $\alpha$ matrix entering our equation
\begin{eqnarray}
&&|\alpha_{1,0}|^2=|\alpha_{-1,0}|^2=\frac{4\pi^2}{a^2}\frac{\hbar}{2\me\Omega}=\frac{\pi\Phi_0}{\Phi \left(1+\eta^2\right)^{1/2}} \;\;\; \textrm{and}\nonumber\\
&&|\alpha_{0,1}|^2=|\alpha_{0,-1}|^2=\frac{\pi \Phi_0}{\Phi\left(1+\eta^2\right)^{3/2}}.
\end{eqnarray}
To obtain the above results we used the definition for $\mu$ and $\Omega$ and the coupling constant $\eta$ given in Eqs.~(\ref{mass polaritonic parameters}), (\ref{upper polariton frequency}) and (\ref{fractal coupling}) respectively. In addition, we use the definitions for $m_p$, $M$, $\Omega$, and $A^{k_x}_n$ given respectively in Eqs.~(\ref{mass polaritonic parameters}),~(\ref{upper polariton frequency}) and~(\ref{HO-shift}) and we find
\begin{equation}
\frac{m_p}{\sqrt{2}M\omega_c}G^y_{\pm 1}A^{k_x}_n=\frac{\pm 1}{1+\eta^2}\frac{2\pi\Phi_0}{\Phi}\left(\frac{ak_x}{2\pi}+n\right).
\end{equation}
After these manipulations we obtain
\begin{widetext}
\begin{equation}\label{Polariton Harper}
\begin{aligned}
&\left[\frac{\hbar^2\left(\sqrt{2}\omega_c k_w+G^y_{m}\right)^2}{2\me(1+\eta^{-2})}+\mathcal{E}_0-E_{k_x,k_w}\right]U^{k_x,k_w}_{n,m}+ t_1(\Phi,\eta)\left(U^{k_x,k_w}_{n-1,m}+U^{k_x,k_w}_{n+1,m}\right)\\
&+t_2(\Phi,\eta)\left[U^{k_x,k_w}_{n,m-1}\exp\left(\frac{-\textrm{i}2\pi\Phi_0}{\Phi(1+\eta^2)}\left(\frac{ak_x}{2\pi}+n\right)\right)+U^{k_x,k_w}_{n,m+1}\exp\left(\frac{\textrm{i}2\pi\Phi_0}{\Phi(1+\eta^2)}\left(\frac{ak_x}{2\pi}+n\right)\right)\right]=0,
\end{aligned}
\end{equation}
\end{widetext}
where we defined the functions $t_1(\Phi,\eta)$ and $t_2(\Phi,\eta)$, shown below, which play a similar role as the hopping matrix elements in a tight-binding description. 
\begin{eqnarray}\label{polariton hoppings}
t_{1}(\Phi,\eta)&=&V_0 \exp\left(-\frac{\pi\Phi_0}{2\Phi(1+\eta^2)^{1/2}}\right)\nonumber \\ t_2(\Phi,\eta)&=&V_0 \exp\left(-\frac{\pi\Phi_0}{2\Phi(1+\eta^2)^{3/2}}\right)
\end{eqnarray}
The hopping functions above depend on the relative magnetic flux and the light-matter coupling $\eta$. Equation~(\ref{Polariton Harper}) is a polaritonic extension of the Harper equation. This can be understood from the fact that in the limit $\eta\rightarrow 0$ the polaritonic Harper equation~(\ref{Polariton Harper}) reduces to the standard Harper equation~\cite{Harper_1955}. We will see how this limit can be performed and discuss this important point in detail in Section~\ref{Harper Duality}.

However, the polaritonic Harper equation~(\ref{Polariton Harper}) has several important differences to the standard Harper equation. First of all, Eq.~(\ref{Polariton Harper}) does not describe simply electrons on a lattice under the influence of a magnetic field, but it describes Landau polaritons on a lattice. Further, there is an additional degree of freedom $k_w$ corresponding to the polaritonic Bloch wave in the $w$ direction. Most importantly, the polaritonic Harper equation does not only depend parametrically on the relative magnetic flux $\Phi/\Phi_0$, but also on the light-matter coupling constant $\eta=\omega_p/\omega_c$. This opens the possibility of not only having a fractal/self-similar spectrum as a function of the relative flux, but also a fractal as a function of the light-matter coupling constant $\eta$. The coupling constant $\eta$ can be tuned experimentally either by varying the strength of the external magnetic field (i.e., changing the cyclotron frequency $\omega_c$) or by varying the diamagnetic frequency $\omega_p$ via the 2D electron density and the fundamental cavity frequency, or by shaping the cavity environment in order to achieve a smaller effective volume~\cite{paravacini2019, FaistCavityHall}.

To test the existence of this polaritonic fractal, we plot the energy spectrum of the polaritonic Harper equation~(\ref{Polariton Harper}) as a function of the light-matter coupling $\eta$ for different values of the relative magnetic flux $\Phi/\Phi_0$. We note that in all the computations performed in this and the following section the momenta $k_x$ and $k_w$ are taken to be equal to zero. This is done for computational simplicity and because we found that the inclusion of different momenta throughout the Brillouin zone has very little influence on the energy spectra and fractal patterns.

First, we start with computing the energy spectrum as a function of $\eta$ for a relatively small magnetic flux $\Phi/\Phi_0=0.1$. In Fig.~\ref{Fractal_pol_0,1} the energy spectrum consists of well separated energy levels without much overlap between them. For small $\eta$ the gaps between the energy levels are small and as the light-matter coupling $\eta$ increases the energies fan out and the gaps increase without any significant pattern emerging.

Subsequently, we double the magnetic flux and show in Fig.~\ref{Fractal_pol_0,2} the respective energy spectrum. For $\Phi/\Phi_0=0.2$ we see that the energies broaden and each energy band demonstrates an internal oscillatory behavior as a function of $\eta$. However, the energy bands are still well-separated and there is not much overlap between them.

In Fig.~\ref{Fractal_pol_1} we increase even further the relative magnetic flux and we plot the polaritonic energies for magnetic flux equal to the flux quantum, $\Phi/\Phi_0=1$, and varying $\eta$. In this case we see that a self-similar pattern emerges as a function of the light-matter coupling, which is similar to well-known fractal pattern of the Hofstadter butterfly~\cite{Hofstadter}.

In addition, we compute and plot in Fig.~\ref{Fractal_pol_g5} the polaritonic energy spectrum for $\Phi/\Phi_0=1$ but now for $\eta$ ranging from $0$ to $5$. With respect to $\eta=1$ we see that on the left and right there is self-similarity but clearly the pattern is different on the two sides of the plot. 

From these computations of the energy spectrum for different magnetic fluxes and over different regimes of light-matter interaction we conclude that for 2D periodic materials strongly coupled to the quantized cavity field and placed perpendicular to a homogeneous magnetic field there is not only a fractal spectrum emerging as a function of the magnetic flux but there is also a novel fractal pattern showing up as a function of the light-matter coupling $\eta$. This implies that fractal structures do not only appear due to the magnetic field but also due to the quantized cavity field and the interaction of the Landau polariton states with the periodic potential of the material. Thus, what we have presented here introduces the novel concept of polaritonic fractals or \textit{fractal polaritons}. To the best of our knowledge such a phenomenon has not been reported before.

\subsection{Cavity Fluctuations in the Hofstadter Butterfly}\label{Cavity Fluctuations in Hofstadter}

\begin{figure}
     \begin{subfigure}[b]{0.5\textwidth}
        \includegraphics[height=6.3cm,width=\linewidth]{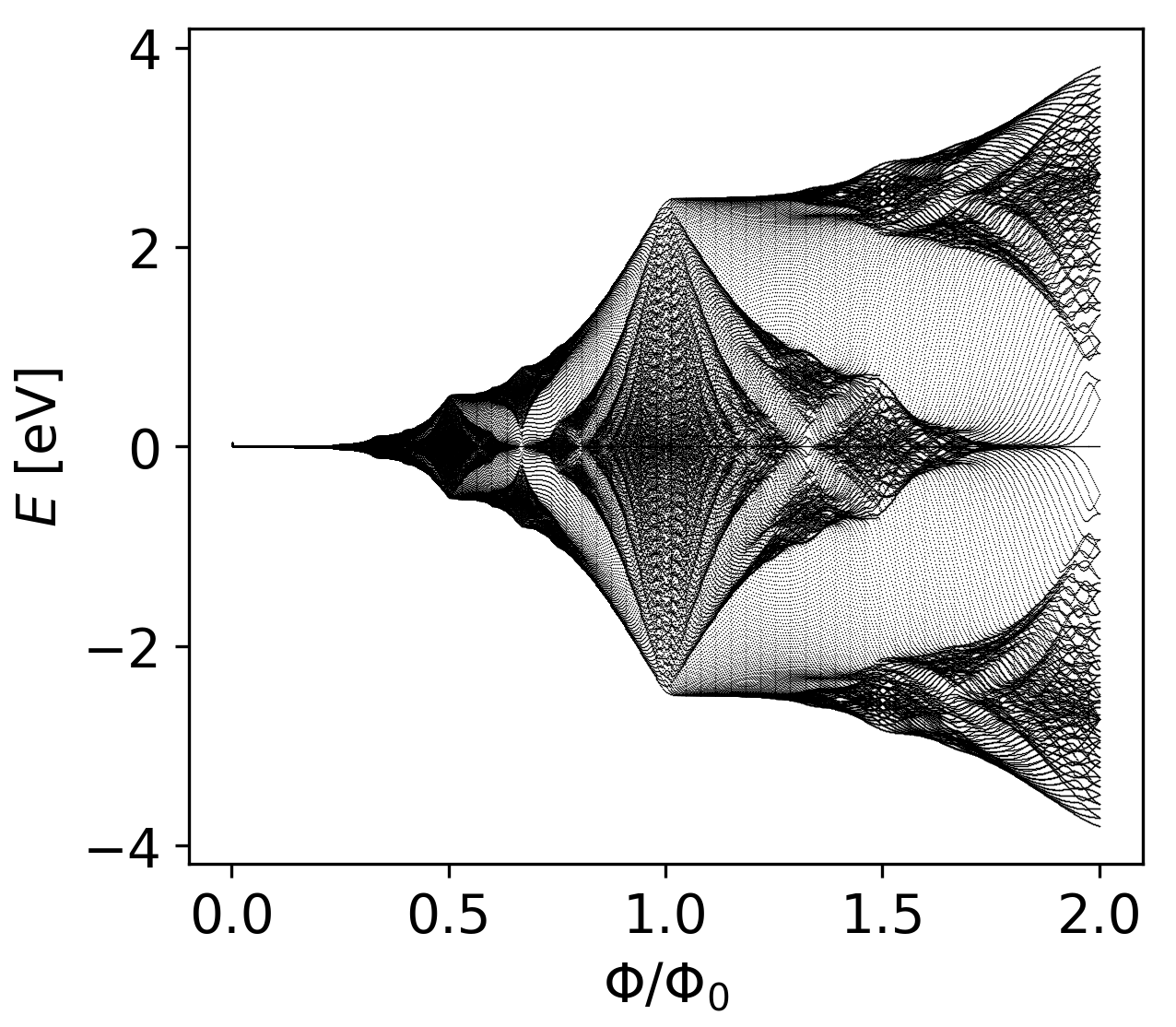}
 \caption{\label{Hofstadter fluctuations small}Energy spectrum as a function of the relative magnetic flux $\Phi/\Phi_0$ for $\omega_p=10^{-3}\textrm{THz}$ (weak coupling).  }
     \end{subfigure}
     \hfill
     \begin{subfigure}[b]{0.5\textwidth}
         \includegraphics[height=6.3cm,width=\linewidth]{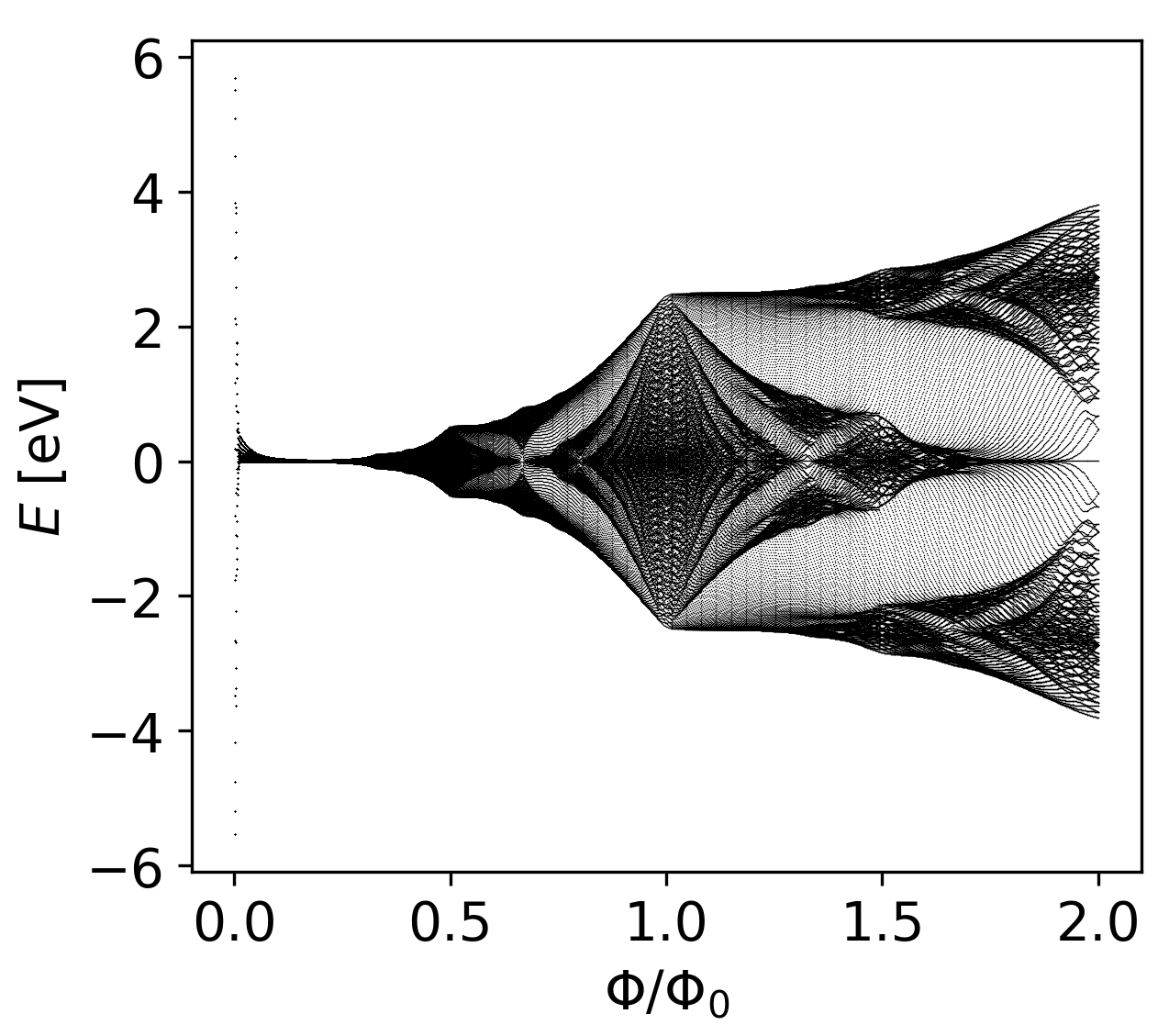}
 \caption{\label{Hofstadter fluctuations medium}Energy spectrum as a function of the relative magnetic flux $\Phi/\Phi_0$ for $\omega_p=10^{-1}\textrm{THz}$ (intermediate coupling).  }
 \end{subfigure}
     \hfill
      \begin{subfigure}[b]{0.5\textwidth}
     \hfill
      \includegraphics[height=6.3cm,width=\linewidth]{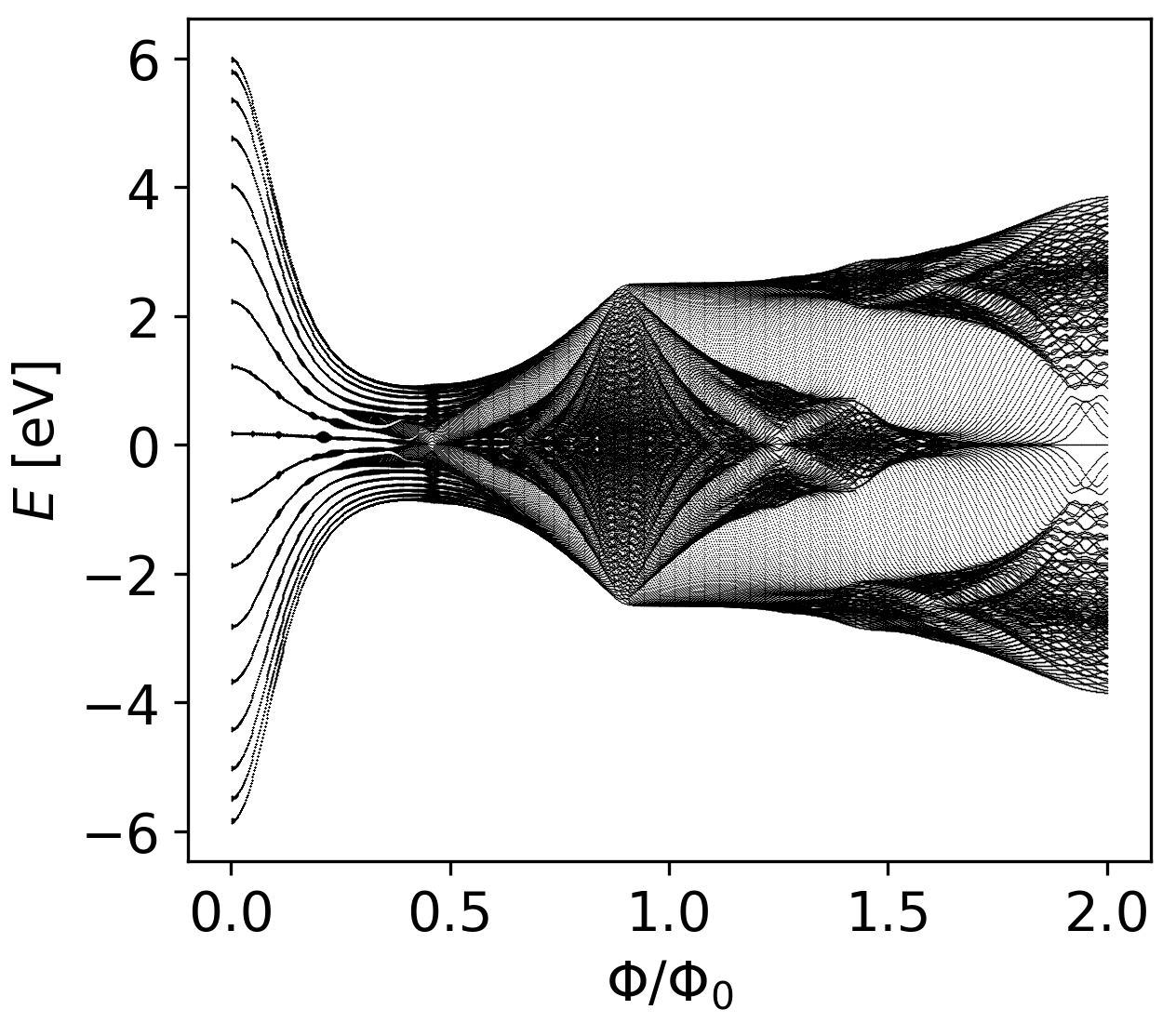}
 \caption{\label{Hofstadter fluctuations large}Energy spectrum as a function of the relative magnetic flux $\Phi/\Phi_0$ for $\omega_p=1\textrm{THz}$ (strong coupling).}
	\end{subfigure}
	   \caption{Energy spectra for the square cosine potential as a function of the relative magnetic flux $\Phi/\Phi_0$ for different values of the diamagnetic frequency $\omega_p$.}
	\label{Butterfly fluctuations}
\end{figure}

So far, we computed the energy spectrum for a fixed value of the relative magnetic flux $\Phi/\Phi_0$ as a function of the light-matter coupling $\eta$ and we demonstrated that a fractal polaritonic butterfly spectrum arises. Now we will do the opposite, namely fix the value for the strength of the quantized photon field determined by the diamagnetic frequency $\omega_p$ (see Eq.~(\ref{effective vector potential})) and then plot the energy spectrum as a function of the relative magnetic flux $\Phi/\Phi_0$, as it was done by Hofstadter for the purely electronic problem~\cite{Hofstadter}. For our computations we will use again the square cosine potential, whose energy spectrum is given by the polaritonic Harper equation~\ref{Polariton Harper}.

First, we compute the energy spectrum for a relatively small value of $\omega_p=10^{-3} \textrm{THz}$, which is a thousand times smaller than the standard terahertz cavity modes used in the setting of Landau polaritons~\cite{paravacini2019, rokaj2019, li2018}. In Fig.~\ref{Hofstadter fluctuations small} we show the energy spectrum as a function of the relative magnetic flux. In the low flux regime we see the energy of the lowest Landau level without any splitting. As the flux increases the energy level broadens and mini-gaps start showing up in the region $0.25 \leq \Phi/\Phi_0\leq 0.5$. For $\Phi/\Phi_0 > 0.5$ the gaps become much larger and the fractal pattern of the Hofstadter butterfly emerges. Due to the small value of $\omega_p$ no modification of the butterfly spectrum is visible and we recover the standard butterfly pattern. Only at $\Phi/\Phi_0 \approx 0$ we see a little spike deviating from the linear Landau-level dispersion, which is due to the fact that for small fluxes the diamagnetic frequency $\omega_p$ is larger than the cyclotron frequency $\omega_c$.

Next, in Fig.~\ref{Hofstadter fluctuations medium} we increase the diamagnetic frequency by two orders of magnitude to $\omega_p=10^{-1}\textrm{THz}$. In this case, around $\Phi/\Phi_0=0$ the energies of the system spread over a wide range and the deviation from the linear dispersion persists over a larger region of magnetic fluxes because the diamagnetic frequency is comparable to $\omega_c$ in a larger region. However, as the magnetic flux increases, we get the linear dispersion of the Landau level which then evolves into the Hofstadter butterfly spectrum just like in Fig.~\ref{Hofstadter fluctuations small}.

Finally, we choose the diamagnetic frequency to be of the order of one terahertz, $\omega_p=1\textrm{THz}$. As we showed in~\cite{rokaj2019}, such a value for $\omega_p$ is within experimental reach in Landau polariton platforms~\cite{paravacini2019, ScalariScience, li2018} where the fundamental cavity frequency $\omega_{\textrm{cav}}$ is also in the terahertz and the 2D electron densities are of the order $n_{\textrm{2D}}\sim 10^{12}\textrm{cm}^{-2}$ . In Fig.~\ref{Hofstadter fluctuations large} we plot the energy spectrum as a function of the relative magnetic flux. In this case we see that in the region $0\leq\Phi/\Phi_0\leq 0.5$ the linear dispersion of the Landau level no longer shows up. In this region the energy spectrum has been completely deformed due to the formation of Landau polariton states as a consequence of the large value of $\omega_p$~\cite{rokaj2019}. The energy spectrum in this regime consists of a set of energy levels with finite width which spread like tentacles over the energies from $-6\textrm{eV}$ to $+6\textrm{eV}$. Then, as the magnetic flux increases the energy levels merge together and recombine to form the fractal spectrum of the Hofstadter butterfly. Figure~\ref{Hofstadter fluctuations large} shows clearly that the fractal spectrum of the Hofstadter butterfly gets strongly modified by the interaction with the cavity photons. This phenomenon is a novel prediction of QED-Bloch theory.  

\subsection{Polaritonic Butterfly in Hexagonal Lattice}
\begin{figure*}
     \centering
     \begin{subfigure}[b]{0.48\textwidth}
         \centering
        \includegraphics[width= \columnwidth]{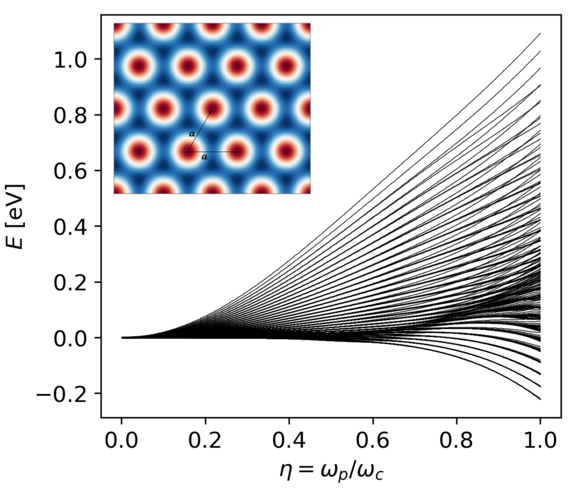}
    \caption{\label{Hex_Polaritons_0,2}Energy spectrum as a function of the light-matter coupling $\eta=\omega_p/\omega_c$ for magnetic flux ratio $\Phi/\Phi_0=0.2$. }
     \end{subfigure}
     \hfill
     \begin{subfigure}[b]{0.48\textwidth}
         \centering
         \includegraphics[width= \columnwidth]{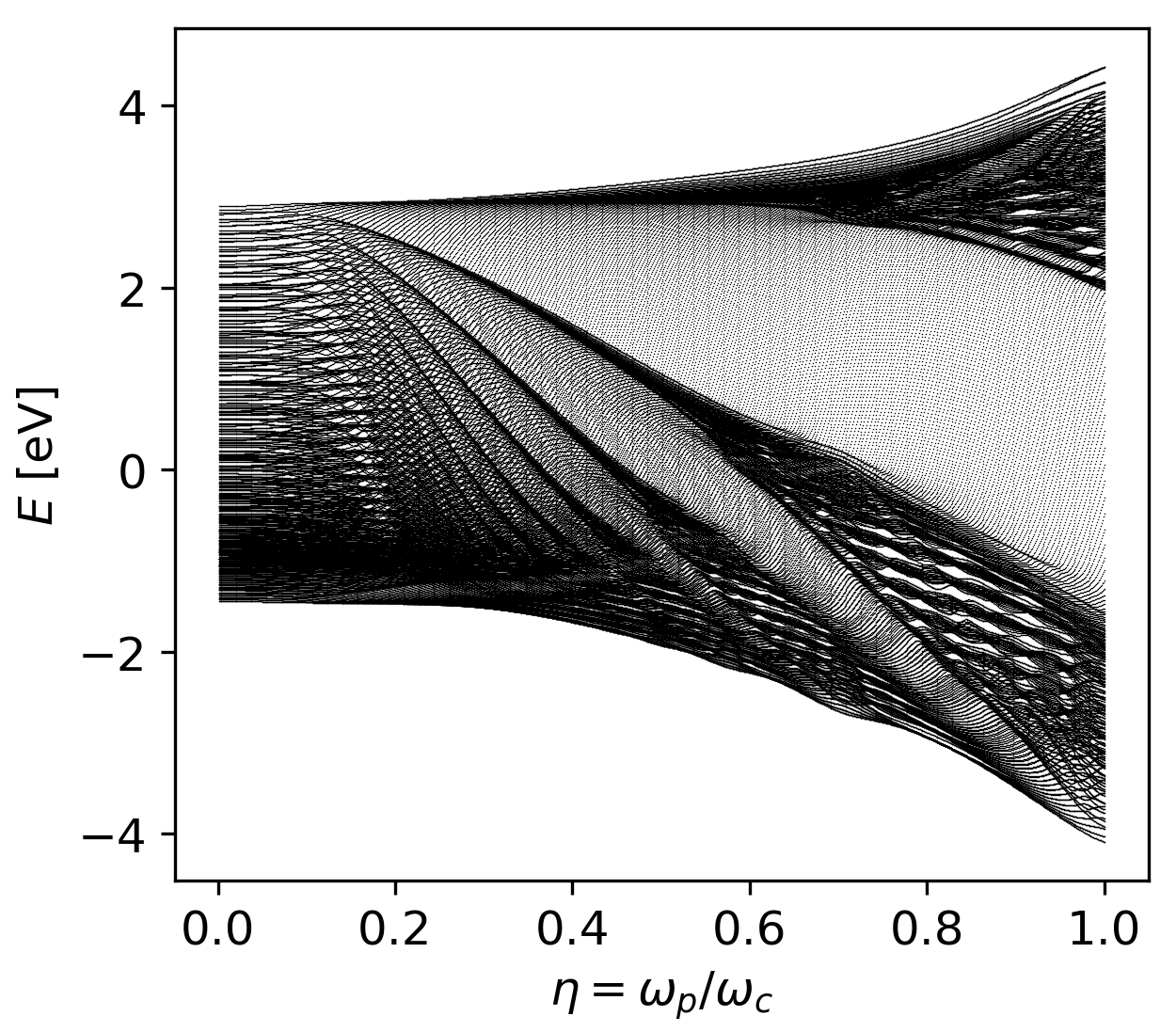}
    \caption{\label{Hex_Polaritons}Energy spectrum as a function of the light-matter coupling $\eta=\omega_p/\omega_c$ for magnetic flux ratio $\Phi/\Phi_0=1$.}
 \end{subfigure}
     \caption{Energy spectra as a function of the light-matter coupling $\eta$ for the hexagonal cosine potential for different values of the relative magnetic flux $\Phi/\Phi_0$. The inset displays the spatial profile of the applied hexagonal cosine lattice potential.}
        \label{Pol Butterfly Hex}
\end{figure*}

In addition to the results that we presented for the square-lattice cosine potential we now also apply the QED-Bloch central equation (\ref{QED-Bloch Central}) for the case of a hexagonal cosine potential (see the inset in Fig.~\ref{Hex_Polaritons_0,2}). We note that in the case of the hexagonal lattice the lattice constants $a_1$ and $a_2$ are equal and we choose them to be the same with the square-lattice case, $a_1=a_2=15\textrm{nm}$. Further, the parameter $V_0$ which controls the strength of the lattice potential is the same as with the square-cosine potential before, $V_0=3\textrm{eV}$.

First, we compute the energy spectrum of the combined electron-photon system as a function of $\eta$, for a relatively small value of the relative magnetic flux $\Phi/\Phi_0=0.2$. In Fig.~\ref{Hex_Polaritons_0,2} we plot the energy spectrum and we see that for small values of $\eta$ the energies levels are well concentrated but as the light-matter coupling increases the energy levels fan out (as in Fig.~\ref{Fractal_pol_0,1} for the square lattice) and the energy gaps increase. However, no significant pattern emerges due to the small value of the magnetic flux. 

Subsequently, we increase the value for the relative magnetic flux to $\Phi/\Phi_0=1$. In Fig.~\ref{Hex_Polaritons} we plot the energy spectrum as a function of the light-matter coupling $\eta$ and we clearly see that a self-similar pattern shows up. In comparison to the energy spectrum for the square lattice depicted in Fig.~\ref{Fractal_pol_1}, we notice that for the hexagonal lattice the energy spectrum it is not vertically symmetric. This is typical for hexagonal lattices and was also reported for the Hofstadter butterfly in the hexagonal lattice case~\cite{ClaroWannier}.

\subsection{Connection to Floquet Engineering}

In the recent years, another topic of interest in the field of 2D materials in homogeneous magnetic fields has been the Floquet driving of the Hofstadter butterfly~\cite{FloquetButterflyKagome, FloquetButterflySquare, GenesisFloquetButterfly}. Also the connection between Floquet driving and cavity engineering has attracted considerable attention and has been explored extensively~\cite{SentefCavityFloquet, schafer2018ab, ChiralCavities, EckhardtChain, RoadmapReview}. In what follows, we will try to compare some of the results coming from our QED-Bloch theory to the ones obtained with Floquet driving.

A generic feature of Floquet theory is that the Floquet field produces photonic copies, above and below the original bare electronic bands~\cite{ChiralCavities, RoadmapReview}. The photonic Floquet copies show up because the Floquet field generates states in the time-domain which dress the electronic states. Then, the quasienergy spectrum can become dense, as it was shown in Ref.~\cite{KohnFloquet}.

Such photonic copies are also a feature of QED. However, there is a fundamental difference between non-relativistic QED and Floquet theory. The Floquet Hamiltonian is unbounded from below and produces states which have an arbitrarily negative energy. On the other hand, the Pauli-Fierz Hamiltonian of non-relativistic QED is bounded from below and the photons do not produce states with arbitrarily negative energies~\cite{spohn2004}. This means that electron-photon systems in QED have a stable ground state~\cite{spohn2004, rokaj2017, schaeferquadratic}. This is an advantage of QED compared to semi-classical Floquet theory. Thus, in non-relativistic QED one cannot expect to obtain photonic copies of the energy bands below the original bare electronic bands, yet above such copies occur.  

\begin{figure*}
     \centering
     \begin{subfigure}[b]{0.45\textwidth}
         \centering
        \includegraphics[width=\columnwidth]{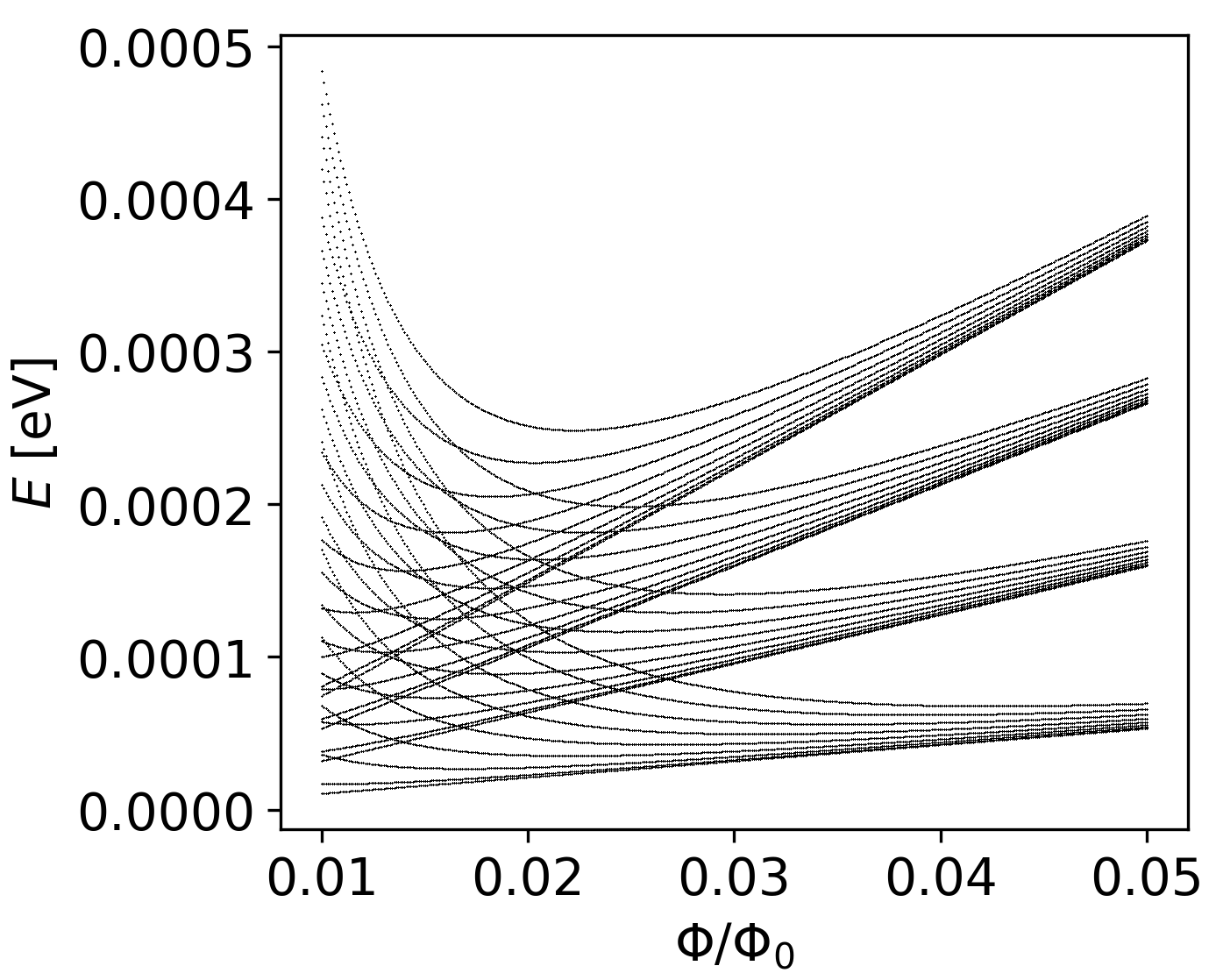}
    \caption{\label{Floquetcopies}Energy spectrum as a function of $\Phi/\Phi_0$ for a small diamagnetic frequency $\omega_p=10^{-3}$THz with four Landau polariton states.}
     \end{subfigure}
     \hfill
     \begin{subfigure}[b]{0.48\textwidth}
         \centering
         \includegraphics[width=0.85\columnwidth]{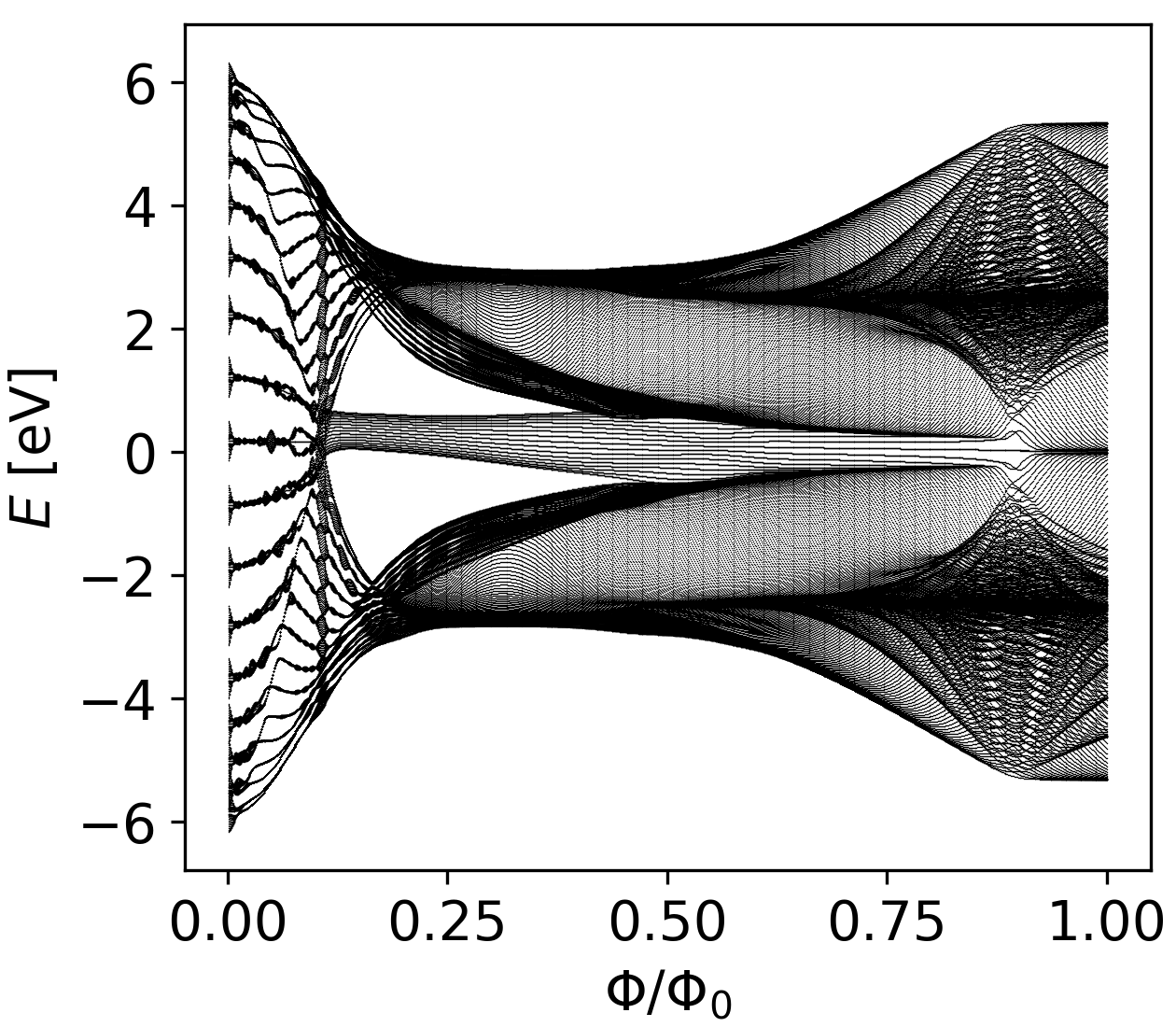}
    \caption{\label{Floquetmixing}Energy spectrum as a function of $\Phi/\Phi_0$ for a large diamagnetic frequency $\omega_p=1$THz with two Landau polariton states.}
 \end{subfigure}
     \caption{Energy spectra for the square cosine potential as a function of the relative magnetic flux $\Phi/\Phi_0$ with higher Landau polariton states, for two different values of the diamagnetic frequency $\omega_p$ which correspond to weak (left) and strong (right) coupling to the cavity photons.}
        \label{Polariton Floquet}
\end{figure*}

To demonstrate these two basic features, (i) the photonic copies and (ii) denseness of the spectrum, we plot in Fig.\ref{Floquetcopies}  the energy spectrum for the square lattice, in the small flux regime and for small diamagnetic frequency $\omega_p=10^{-3}$THz with four Landau polariton states included. As it is shown in Fig.~\ref{Floquetcopies}, we have indeed the higher states introducing copies of the lowest polariton band. The copies here are well-separated without much interference, and the energy separation is given by the Landau polariton excitation $\hbar\Omega$. It is important to note that we do not work with a tensor-product basis between the electrons and photons, like in Floquet theory, and as consequence the higher states are polaritonic and the energy separation is the one of the Landau polariton. In addition, we clearly see that the energy spectrum in the small flux regime becomes dense, as expected~\cite{KohnFloquet}. Thus, our QED-Bloch theory recovers these two basic features of Floquet theory. It is important to note that for $\omega_p \ll 10^{-3}$THz the states between Landau polaritons which produce the dense spectrum do not show up, because we practically have no photonic contribution, and the spectrum is no longer dense, as expected in the case of no photons.

Furthermore, in the studies of the Hofstadter butterfly under Floquet driving~\cite{FloquetButterflyKagome, FloquetButterflySquare, GenesisFloquetButterfly}, another interesting phenomenon has been identified: In the regime where the frequency of the driving field is small, such that the energy of the photon is not much larger than the band width, the Floquet copies overlap with the original bare electronic Hofstadter butterfly and modify the spectrum substantially~\cite{FloquetButterflyKagome, FloquetButterflySquare, GenesisFloquetButterfly}.

This regime, from the perspective of cavity QED can be understood as a strong-coupling scenario where the electronic and the photonic states mix strongly. To compare our QED-Bloch theory to the Floquet results we choose the diamagnetic frequency to be large $\omega_p=1$THz and we plot the energy spectrum as a function of $\Phi/\Phi_0$ for the square cosine potential with two Landau polariton states in Fig.~\ref{Floquetmixing}. As previously, the photon field introduces a copy of the Hofstadter butterfly, but in this case the copies overlap and interfere. This leads to a substantial modification of the result we obtained in the case where only the lowest Landau polariton was taken into account (see Fig.~\ref{Hofstadter fluctuations large}). As it is shown in Fig.~\ref{Floquetmixing}, the two butterflies merge together and form a new fractal pattern. This result is in qualitative agreement with the Floquet driving of the Hofstadter butterfly. To conclude, the results that we presented in this subsection show that our QED-Bloch theory recovers the common features of Floquet theory and of the Floquet driving of the Hofstadter butterfly. 

\section{Harper Equation \& Minimal Coupling-Peierls Phase Duality}\label{Harper Duality}

Our aim in this section is to connect the polaritonic Harper equation that we derived in Eq.~(\ref{Polariton Harper}) to the well-known Harper equation~\cite{Harper_1955} and the fractal spectrum of the Hofstadter butterfly~\cite{Hofstadter}. Both, the Harper equation and the butterfly spectrum were derived from a one-band tight-binding model with next-neighbor hopping in real space on a square lattice, with the electrons coupled to the magnetic field via the Peierls substitution. We note that using the Peierls substitution relies on the assumption that the external magnetic field is weak enough such that no mixing between different bands occurs~\cite{Harper_1955, LuttingerTBPeierls}.

The polaritonic Harper equation we derived is for a square lattice potential as well, and we chose the cosine potential because it introduces only next-neighbor hopping in Fourier space, as the only non-zero Fourier components of the potential are $V_{\pm1,0}=V_{0,\pm1}=V_0$. With the extra assumption of only the lowest Landau polariton being occupied we obtained Eq.~(\ref{Polariton Harper}). To connect now the polaritonic Harper equation to the original Harper equation we need to take the limit of the light-matter coupling to zero, $\eta\rightarrow 0$. Then the kinetic term which depends on the polaritonic momentum $k_w$ vanishes because $\eta^{-2}$ goes to infinity and the equation becomes completely independent of $k_w$ and as a consequence also the Fourier index $m$ can be dropped. In addition, for $\eta=0$, the hopping functions $t_1(\Phi,\eta)$ and $t_2(\Phi,\eta)$ defined in Eq.~(\ref{polariton hoppings}) become equal,
\begin{equation}\label{flux hopping}
t_1(\Phi,\eta)=t_2(\Phi,\eta)=V_0\rme^{-\frac{\pi\Phi_0}{2\Phi}}\equiv t(\Phi).
\end{equation}
Also, in the described limit the energy of the lowest Landau polariton $\mathcal{E}_0=\hbar\Omega/2$ for zero light-matter coupling becomes equal to the energy of the lowest Landau level $\mathcal{E}_0=\hbar\omega_c/2$. This means that for $\eta\rightarrow 0$ the polaritonic Harper equation does not describe Landau polaritons on a lattice but actually Landau levels on a lattice and it takes the form
\begin{widetext}
\begin{equation}\label{Unscaled Harper}
U^{k_x}_{n}\left(\frac{\hbar\omega_c}{2}-E_{k_x}\right) + t(\Phi)\left[U^{k_x}_{n-1}+U^{k_x}_{n+1}+2U^{k_x}_n \cos\left(\frac{2\pi \Phi_0}{\Phi} \left(\frac{ak_x}{2\pi}+n\right)\right)\right]=0.
\end{equation}
\end{widetext}
In the work of Hofstadter~\cite{Hofstadter} the fractal spectrum appears not for the energy itself, $E$, but for the unitless scaled energy, $\mathcal{E}=E/t$, where $E$ is divided by the constant hopping parameter $t$. Of course this does not make a difference within the tight-binding model because the hopping parameter $t$ is a constant anyway. On the contrary, for the minimal-coupling Hamiltonian the magnetic field is part of the covariant (physical) momentum of the electron. Thus, the kinetic energy of the electrons naturally depends on the magnetic field and as a consequence the hopping (which represents the kinetic energy in the tight-binding approach) should be a function of the magnetic field as well. In our setting we have the flux-dependent hopping parameter $t(\Phi)$ defined in Eq.~(\ref{flux hopping}) and as a consequence the dimensionless scaled energies can be defined as
\begin{equation}\label{scaled energies}
\mathcal{E}_{k_x}=\frac{1}{t(\Phi)}\left(E_{k_x}-\frac{\hbar\omega_c}{2}\right)=\frac{\rme^{\frac{\pi\Phi_0}{2\Phi}}}{V_0}\left(E_{k_x}-\frac{\hbar\omega_c}{2}\right).
\end{equation}
With this definition, we find the following equation for the scaled, dimensionless energies of the system,
\begin{equation}\label{Harper equation}
\mathcal{E}_{k_x}U^{k_x}_{n}= U^{k_x}_{n-1}+U^{k_x}_{n+1}+2U^{k_x}_n \cos\left(\frac{2\pi \Phi_0}{\Phi} \left(\frac{ak_x}{2\pi}+n\right)\right).
\end{equation}

\begin{figure*}
     \centering
     \begin{subfigure}[b]{0.48\textwidth}
         \centering
        \includegraphics[width=\linewidth]{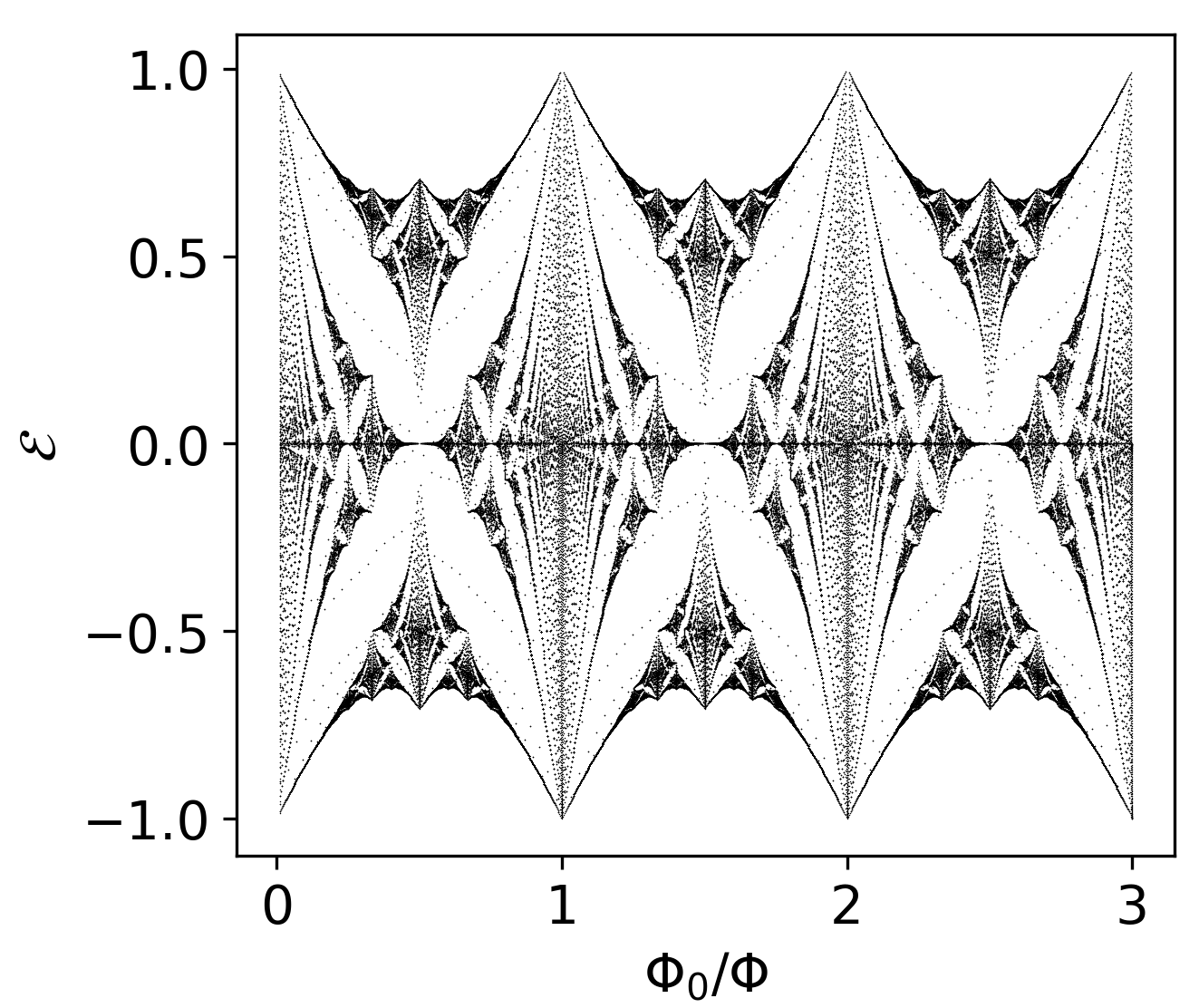}
\caption{\label{Reciprocal Butterfly}Scaled dimensionless energy spectrum obtained from the Harper equation~(\ref{Harper equation}) as a function of the reciprocal magnetic flux $\Phi_0/\Phi$.}
     \end{subfigure}
     \hfill
     \begin{subfigure}[b]{0.48\textwidth}
         \centering
         \includegraphics[width=\linewidth]{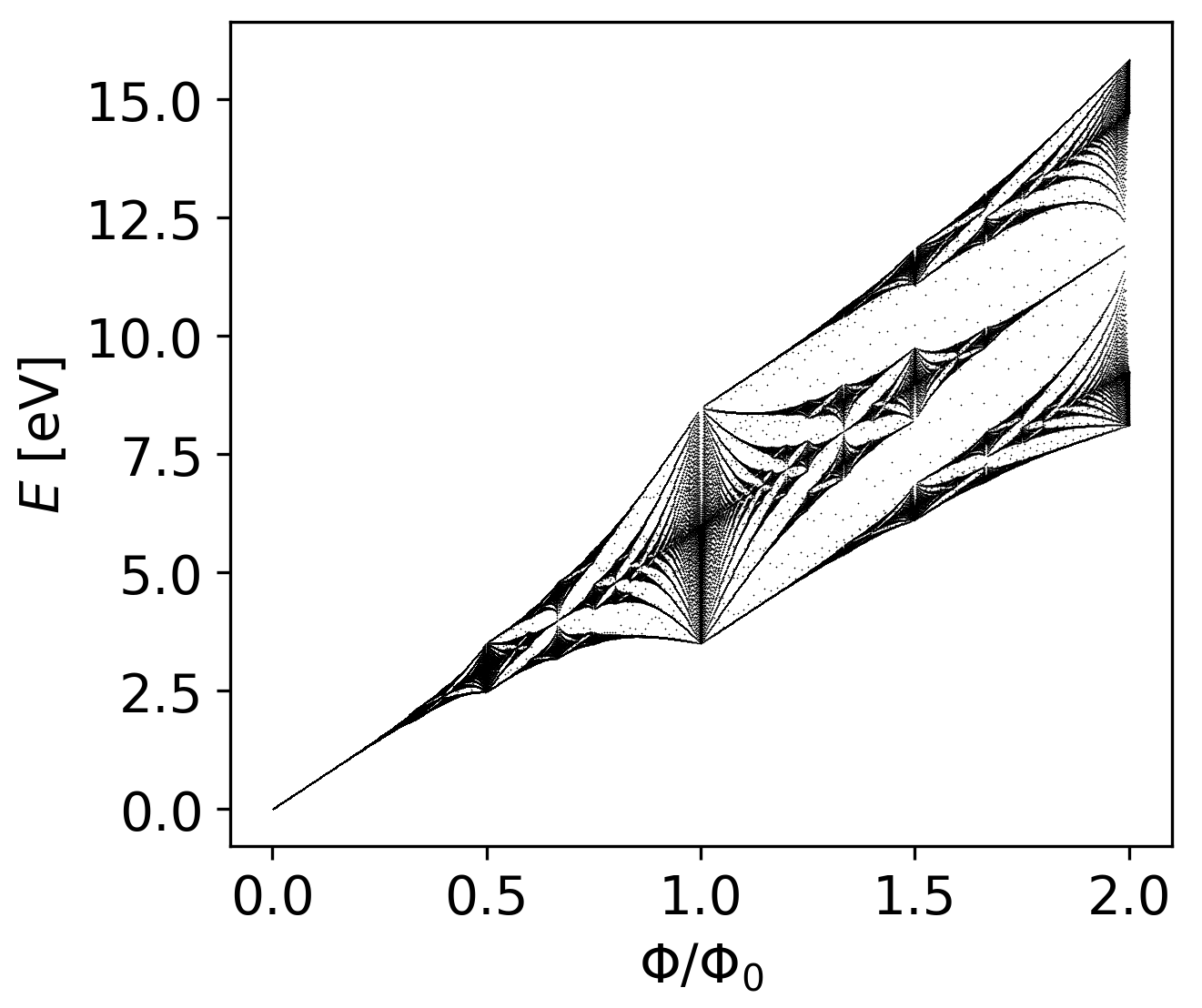}
\caption{\label{Unscaled Butterfly}Energy spectrum as function of the relative magnetic flux $\Phi/\Phi_0$ as given by Eq.~(\ref{Unscaled Harper}). }
 \end{subfigure}
     \caption{Comparison between the scaled and the unscaled spectrum for the square cosine potential. The potential strength is $V_0=3$eV and the lattice constant $a=2$\AA. We use a small lattice constant to make the linear dispersion of the Landau level in Fig.~\ref{Unscaled Butterfly} clearly visible.}
        \label{Scaled and Unscaled Spectra}
\end{figure*}

The equation above is precisely the usual Harper equation~\cite{Harper_1955} and plotting the eigenenergies of this equation we obtain the fractal spectrum of the Hofstadter butterfly~\cite{Hofstadter}, which is depicted in Fig~\ref{Reciprocal Butterfly}. However, there is one important difference: In the original Harper equation the energy spectrum is a function of the relative magnetic flux $\Phi/\Phi_0$ and the corresponding butterfly spectrum also appears as a function of the relative flux $\Phi/\Phi_0$. In our case the energies are a function of the reciprocal relative flux $\Phi_0/\Phi$ and the fractal spectrum appears with respect to the reciprocal flux $\Phi_0/\Phi$. This fact, that starting from the minimal-coupling Hamiltonian one obtains the Hofstadter butterfly as a function of the reciprocal flux $\Phi_0/\Phi$ comprises a fundamental difference to tight-binding models with the Peierls phase, as has been shown and discussed in several publications~\cite{PfannkucheButterfly, Langbein1969, Rauh1975, GeiselButterflyChaos}.

Already Langbein~\cite{Langbein1969} discussed and compared the two approaches while it was noted by Wannier~\cite{Wannier} that it is highly surprising for the energy spectrum to be periodic either in the relative magnetic flux $\Phi/\Phi_0$ or in the reciprocal magnetic flux $\Phi_0/\Phi$, because the minimal-coupling Hamiltonian has a linear and a quadratic dependence with respect to the magnetic field. Both periodic behaviors are in fact an artefact. As it was pointed out by Wannier~\cite{Wannier} this periodicity cannot be physical, and the tight-binding model with the Peierls phase cannot be trusted for magnetic fields beyond one flux quantum, because for such strong magnetic fields the original assumption that each band can be treated separately and that there is no mixing between them will be invalid.

In the case of the minimal-coupling Hamiltonian the exact periodicity is due to the redefinition (the scaling) of the energies in Eq.~(\ref{scaled energies}). This redefinition of the energy spectrum cuts out the linear dependence of the energy of the lowest Landau level on $\omega_c=eB/\me$ and the exponential increase due to the flux dependent hopping parameter $t(\Phi)$. To understand what the actual dependence of the energy spectrum is, we plot the unscaled, dimensionful energies of our system given by Eq.~(\ref{Unscaled Harper}) in Fig.~\ref{Unscaled Butterfly}.

We see that for small fluxes we have the linear dispersion coming from the energy of the lowest Landau level. As the magnetic field increases, the Landau level starts to split and gaps show up. Then, for $\Phi/\Phi_0>1/2$ the fractal nature of the spectrum shows up and the Hofstadter butterfly becomes clearly visible. The Hofstadter butterfly however is not periodic, but it actually spreads out due to the flux-dependent hopping $t(\Phi)$.  Figure~\ref{Unscaled Butterfly} reconciles the two fundamental properties of the minimal-coupling Hamiltonian: (i) the energy has to increase as a function of the magnetic field and (ii) due to the reduced symmetry manifest in the magnetic translation group a splitting of the energy bands for every fractional value of the relative magnetic flux $\Phi/\Phi_0=p/q$ needs to occur, which subsequently leads to the formation of the fractal~\cite{BrownMTG, MTG_I, MTG_II}.

\textit{Dual Descriptions.}---The question that finally arises is: Why do the energy spectra of the minimal-coupling Hamiltonian and the tight-binding model with the Peierls phase differ in such a fundamental way? 

In many cases, these two descriptions for electrons in periodic structures coupled to electromagnetic fields are indeed equivalent descriptions and match at least to some certain accuracy. Typically, this is true for electromagnetic fields slowly varying within the unit cell of the solid. However, the problem with our particular system is that although the magnetic field is constant, the vector potential that actually couples to the electrons is linear in space $\bi{A}_{\textrm{ext}}=-\bi{e}_xBy$. As it was pointed out by Luttinger, the spatial variation of the vector potential makes the Peierls substitution questionable for large magnetic fields~\cite{LuttingerTBPeierls}. This was one of the first papers deriving the tight-binding model with Peierls phase starting from the minimal-coupling Hamiltonian. To arrive at this model, Luttinger had to drop a term from the Hamiltonian which consequently breaks the actual relation between the minimal-coupling Hamiltonian and the tight-binding model with the Peierls phase. 

However, this does not mean that the two descriptions are completely disconnected. They both yield the Harper equation and the Hofstadter butterfly, with the difference that in the one case it shows up as a function of the magnetic flux $\Phi/\Phi_0$ while in the other as a function of the reciprocal flux $\Phi_0/\Phi$. This means that the minimal-coupling Hamiltonian and the tight-binding model with the Peierls phase are not equivalent but they are actually dual. It is important to mention that this duality holds only in the lowest Landau level and for a single band, respectively.

\begin{figure}[H]
\begin{center}
    \includegraphics[height=4cm, width=\linewidth]{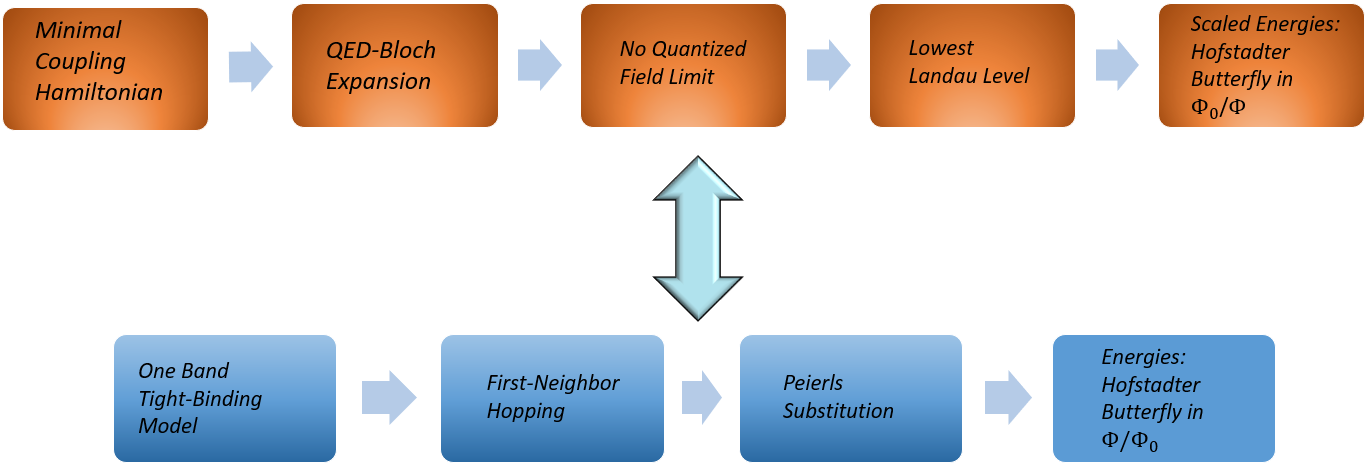}
\caption{\label{Butterfly Duality}Schematic illustration of the duality between the minimal-coupling Hamiltonian and the tight-binding model with Peierls substitution. In both models the Hofstadter butterfly emerges but in a dual fashion. Namely, in the one case as a function of the reciprocal flux while in the other as a function of flux.}
\end{center}
\end{figure}

This dual (or reciprocal) relation between the two models can be understood on the one hand from the fact that in the tight-binding model we have electrons with next-neighbor hopping on a lattice in real space, while in the minimal-coupling Hamiltonian we have next-neighbor hopping in $\textbf{k}$-space. Further, the dual relation manifests itself as a weak-to-strong duality with respect to the strength of the magnetic field, in the sense that the tight-binding model with the Peierls phase mainly describes the regime of not too large fluxes and the butterfly appears as a function of the flux, where the Peierls substitution is still applicable. Instead the minimal-coupling Hamiltonian, restricted in the lowest Landau level, works best for not too small fluxes and the butterfly shows up as a function of the inverse flux. The two approaches match in the regime where $\Phi/\Phi_0 \approx \Phi_0/\Phi$ which is the region around one flux quantum. We note that by the inclusion of higher Landau levels also small magnitudes of the magnetic flux can be correctly described within the minimal-coupling framework. The duality between the two approaches and the steps to obtain the respective Hofstadter butterflies are summarized in Fig.~\ref{Butterfly Duality}. 

\begin{figure}[H]
\begin{center}
    \includegraphics[width=\linewidth]{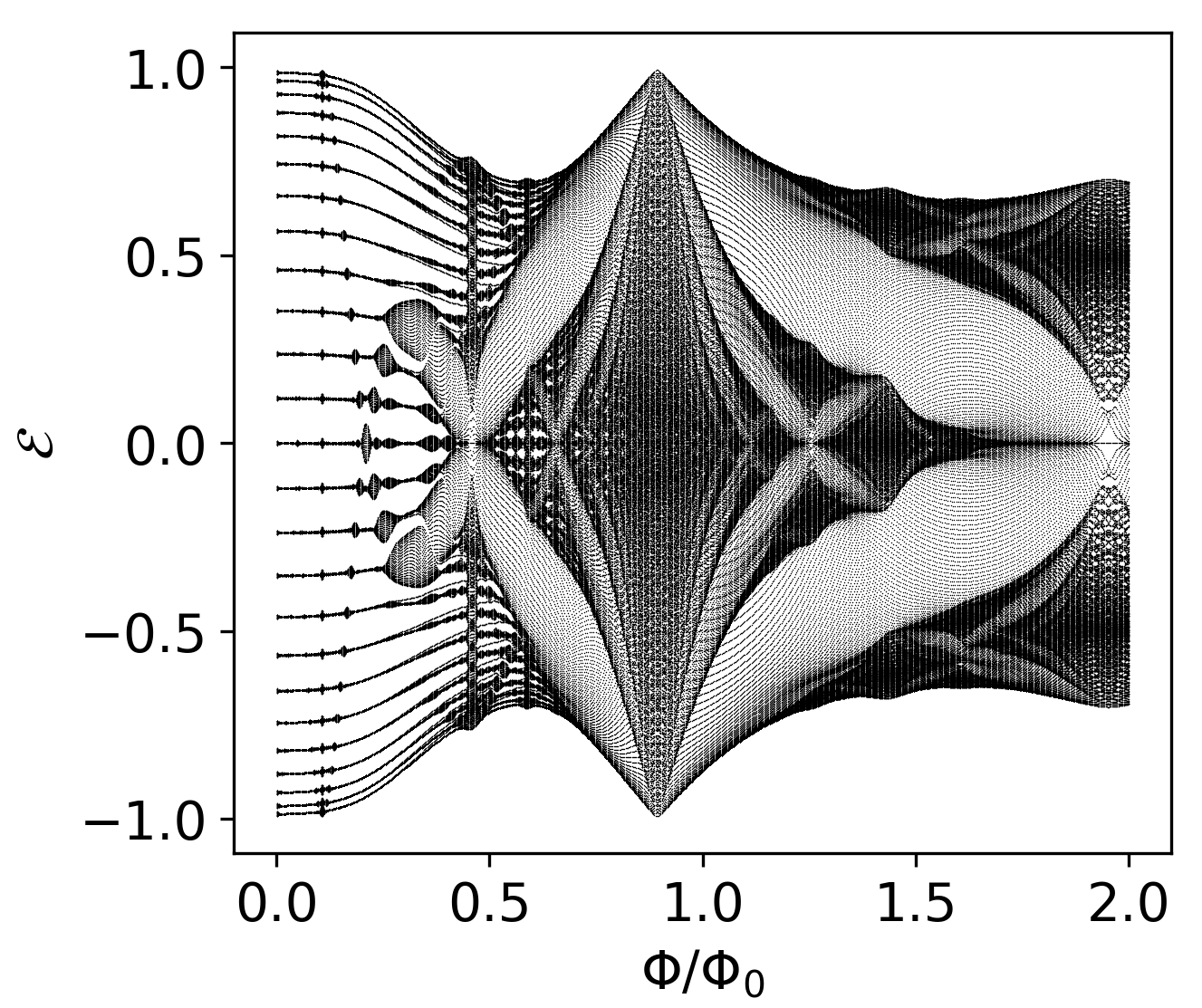}
\caption{\label{Modified TB Butterfly}Scaled dimensionless energy spectrum for the square cosine lattice potential as a function of the magnetic flux $\Phi/\Phi_0$ for $\omega_p=1\textrm{THz}$. The unscaled physical spectrum is shown in Fig.~\ref{Hofstadter fluctuations large}.}
\end{center}
\end{figure}

Having established this duality between the minimal-coupling Hamiltonian and the tight-binding model with the Peierls phase in the semi-classical setting, we would now like to make a connection to the tight-binding model with Peierls phase in the case where the electrons are coupled also to the quantized cavity field. In subsection~\ref{Cavity Fluctuations in Hofstadter} in Fig.~\ref{Hofstadter fluctuations large} we showed how the butterfly pattern coming from the minimal-coupling Hamiltonian gets modified due to strong coupling to a terahertz cavity. In the semi-classical case, to connect the two approaches we applied the scaling transformation defined in Eq.~(\ref{scaled energies}) to the energies of the minimal-coupling Hamiltonian. In the light-matter setting the analogous scaling transformation is
\begin{equation}
    \mathcal{E}_{k_x,k_w}=\frac{E_{k_x,k_w}-\hbar\Omega/2}{t_1(\Phi,\eta)+t_2(\Phi,\eta)},
\end{equation}
where the hopping functions $t_1(\Phi,\eta)$ and $t_2(\Phi,\eta)$ were defined in Eq.~(\ref{polariton hoppings}) and $\hbar\Omega/2$ is the lowest Landau polariton energy. Using this scaling transformation on the polaritonic energies of Fig.~\ref{Hofstadter fluctuations large} we obtain the scaled energy spectrum depicted in Fig.~\ref{Modified TB Butterfly}. The scaled energy spectrum provides a hint on how the original Hofstadter butterfly (coming from the tight-binding model) might look when the 2D material is coupled also to the cavity photon-field in the terahertz regime. In the region of small and intermediate fluxes, where the cavity field dominates, we see that the butterfly pattern is dissolved into distinct energy bands with an internal oscillatory behavior. As the flux becomes larger than half of a flux quantum we see that the butterfly pattern emerges. Still at values around one flux quantum there are modifications of the spectra and the effect of the cavity is noticeable. There is a clear similarity between the butterfly patterns in and outside the cavity around one, but in contrast to the perfectly periodic spectrum (of the Hofstadter butterfly coming from the tight-binding model) the cavity induces a significant distortion that should be visible experimentally.

\section{Cavity Modification of the Integer Hall Conductance}\label{Modification Hall Conductance}

QED-Bloch theory is also applicable to the case of no external potential, which for a homogeneous magnetic field (without a quantized field) constitutes the usual setting of non-interacting Landau levels that describe the integer quantum Hall effect~\cite{Klitzing, LaughlinPRB}.

\begin{figure}[H]
\begin{center}
  \includegraphics[width=0.7\linewidth]{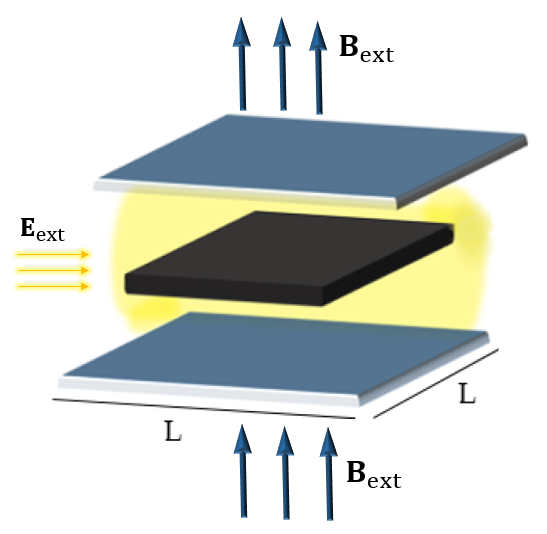}
\caption{\label{Landau Levels Cavity}Cartoon depiction of a 2D electron gas (material in black) confined inside a cavity. The whole system is placed perpendicular to a classical homogeneous magnetic field $\bi{B}_{\textrm{ext}}$, and an external constant electric field $\bi{E}_{\textrm{ext}}$ is applied to the 2D material. }  
\end{center}
\end{figure}

The aim of this section is to study non-interacting Landau levels under strong coupling to the cavity field, in the presence of an external classical electric field (as shown in Fig.~\ref{Landau Levels Cavity}), and to demonstrate that the cavity field modifies the plateaus of the Hall conductance in the integer regime. Such a cavity modification has also been measured recently experimentally~\cite{FaistCavityHall}.

As it was shown in Ref.~\cite{rokaj2019}, in the case where the external potential is zero, $v_{\textrm{ext}}(\bi{r})=0$, the effective Hamiltonian
\begin{equation} 
\Heff=\frac{1}{2\me}\left(\mathrm{i}\hbar \mathbf{\nabla}+e\hat{\bi{\mathcal{A}}}+e\mathbf{A}_{\textrm{ext}}(\bi{r})\right)^2-\frac{\hbar\omega_p}{2}\frac{\partial^2}{\partial u^2}
\end{equation}
is analytically solvable. As a reminder for the reader $\bi{A}_{\textrm{ext}}(\bi{r})=-\bi{e}_xBy$ is the vector potential describing the external magnetic field and $\hat{\bi{\mathcal{A}}}$ is the quantized vector potential of the cavity defined in Eq.~(\ref{effective vector potential}) 

The energy spectrum of this system can be directly obtained from the QED-Bloch central equation~(\ref{QED-Bloch Central}) by simply setting the external potential to zero, $v_{\textrm{ext}}(\bi{r})=0$. Since there is no external potential, the reciprocal lattice vectors have no role and have to be taken equal to zero. This is done by simply setting $n=m=0$ and we get
\begin{equation}
U^{\bi{k}}_{i} \left(\frac{\hbar^2k^2_w}{2M}+\mathcal{E}_i-E_{\bi{k}}\right)=0.
\end{equation}
Then, from the above equation it is clear that the eigenspectrum of the 2D Landau levels coupled to the cavity is 
\begin{equation}\label{Landau polaritons}
E_{\bi{k},i}=\frac{\hbar^2k^2_w}{2M}+\hbar\Omega\left(i+\frac{1}{2}\right).
\end{equation}
Further, the components of $U^{\bi{k}}_{i}$ from the QED-Bloch ansatz defined in Eq.~(\ref{BlochAnsatzHermite}) become trivial and we obtain the full set of eigenfuctions corresponding to the 2D Landau levels in the cavity, 
\begin{equation}\label{Landau polariton States}
\Psi_{\mathbf{k},i}(\mathbf{r}_{w},v)=\rme^{\mathrm{i}\mathbf{k}\cdot\mathbf{r}_w}\phi_i\left(v-\frac{\hbar k_x}{\sqrt{2}\me}\right).
\end{equation}
This is the analytic solution for the non-interacting 2D Landau levels in the cavity that was found in Ref.~\cite{rokaj2019}. The eigenfunctions above are plane waves in the directions $x$ and $w$ because in these directions we have translational invariance. The $v$-dependent eigenfunctions are Hermite funcitons~\cite{GriffithsQM} with argument $v-\hbar k_x/\sqrt{2}\me$. The eigenfunctions in Eq.~(\ref{Landau polariton States}), since they are functions of the combined polaritonic coordinates $w$ and $v$ defined in Eq.~(\ref{w and v coordinates}), describe quasiparticles formed between the Landau levels and the photons, which are known as Landau polaritons. Such Landau polariton states have been studied theoretically~\cite{Hagenmuller2010cyclotron} and have been also observed experimentally~\cite{ScalariScience, Keller2020, li2018}. 

\subsection{Integer Hall Conductance in the Cavity}

Having briefly reviewed how the Landau polariton quasiparticles emerge in QED-Bloch theory, our aim is now to study the implications of these polaritonic quasiparticles for the quantum Hall effect. To compute the Hall conductance of the Landau polaritons we need to add an external potential $V_E(\bi{r})=e\phi_E(\bi{r})=-e\bi{E}_{\textrm{ext}}\cdot \bi{r}=-eEy$ to our effective Hamiltonian, which generates an electric field in the $y$ direction, $\bi{E}_{\textrm{ext}}=-\nabla\phi_E(\bi{r})=E\bi{e}_y$. This is what was done by Laughlin~\cite{LaughlinPRB} for the description of the integer quantum Hall effect in terms of non-interacting Landau levels. We note that in his description the spin degrees of freedom of the electrons were neglected and we do the same here. 

Then, with the addition of the external potential $V_E(\bi{r})$, the effective Hamiltonian for the 2D Landau levels coupled to the cavity is
$\hat{H}_{E}=\Heff+V_{E}(\bi{r})=\Heff-eEy$.
In terms of the polaritonic coordinates $w$ and $v$ introduced in Eq.~(\ref{w and v coordinates}) $\hat{H}_{E}$ takes the simple form
\begin{eqnarray}
\hat{H}_E&=&-\frac{\hbar^2}{2M}\frac{\partial^2}{\partial w^2} -\frac{eE}{\sqrt{2}\omega_c}w\\
&-&\frac{\hbar^2}{2\mu}\frac{\partial^2}{\partial v^2} +\frac{\mu \Omega^2}{2}\left(v+\frac{\textrm{i}\hbar}{\sqrt{2}\me} \frac{\partial}{\partial x}\right)^2 +\frac{e Em_p}{\sqrt{2}M\omega_c}v.\nonumber
\end{eqnarray}
The expression above can be easily derived from Eq.~(\ref{Heff compact}) and from the expression for the $y$ coordinate in terms of the polaritonic coordinates $w$ and $v$,
\begin{equation}
y=\frac{w}{\sqrt{2}\omega_c} -\frac{m_pv}{\sqrt{2}M\omega_c}.
\end{equation}
Due to translational invariance it is clear that the eigenfunctions with respect to $x$ are plane waves of the form $f(x)=\rme^{\textrm{i}k_xx}$. Applying $\hat{H}_E$ on $f(x)=\rme^{\textrm{i}k_xx}$ and then dividing by $f(x)$ we obtain
\begin{eqnarray}
\hat{H}_{E}[k_x]&\equiv&\frac{1}{f(x)}\hat{H}_Ef(x)=-\frac{\hbar^2}{2M}\frac{\partial^2}{\partial w^2} -\frac{e E}{\sqrt{2}\omega_c}w\\
&-&\frac{\hbar^2}{2\mu}\frac{\partial^2}{\partial v^2} +\frac{\mu \Omega^2}{2}\left(v-\frac{\hbar k_x}{\sqrt{2}\me} \right)^2 +\frac{e Em_p}{\sqrt{2}M\omega_c}v.\nonumber
\end{eqnarray}
The $w$-dependent part of the Hamiltonian is independent of the electronic momentum $k_x$ and as a consequence it cannot give any contribution to the induced current along the $x$ direction which we are interested in for the computation of the Hall conductance $\sigma_{xy}$. Thus, for our purpose, we can safely eliminate the $w$-dependent terms from the Hamiltonian and remain with
\begin{equation}
\hat{H}_{E}[k_x]=-\frac{\hbar^2}{2\mu}\frac{\partial^2}{\partial v^2} +\frac{\mu \Omega^2}{2}\left(v-\frac{\hbar k_x}{\sqrt{2}\me} \right)^2 +\frac{\sqrt{2}\me E}{B(1+\eta^2)}v.
\end{equation}
In the previous equation we substituted the definitions for the cyclotron frequency $\omega_c=eB/\me$ and the mass parameters $M$ and $m_p$ given by Eq.~(\ref{mass polaritonic parameters}), and we introduced the light-matter coupling $\eta=\omega_p/\omega_c$.
Next we perform a square completion and the Hamiltonian takes the form of a shifted harmonic oscillator
\begin{widetext}
\begin{equation}
\hat{H}_{E}[k_x]=-\frac{\hbar^2}{2\mu}\frac{\partial^2}{\partial v^2} +\frac{\mu \Omega^2}{2}\left(v-\frac{\hbar k_x}{\sqrt{2}\me} +\frac{E}{\sqrt{2}B(1+\eta^2)}\right)^2 +\frac{\hbar k_x E}{B(1+\eta^2)} -\frac{\me}{2}\left(\frac{E}{B(1+\eta^2)}\right)^2.
\end{equation}
\end{widetext}
In order to obtain the above expression we also used the definitions for $\mu$ and $\Omega$ given by Eqs.~(\ref{mass polaritonic parameters}) and~(\ref{upper polariton frequency}). The eigenfunctions of the above Hamiltonian are Hermite functions $\phi_n(Z)$ which depend on the variable
\begin{equation}
Z=v-\frac{\hbar k_x}{\sqrt{2}\me} +\frac{E}{\sqrt{2}B(1+\eta^2)}
\end{equation}
with eigenenergies 
\begin{equation}
E_{n,k_x}= \hbar\Omega\left(n+\frac{1}{2}\right)+\frac{\hbar k_x E}{B(1+\eta^2)} -\frac{\me}{2}\left(\frac{E}{B(1+\eta^2)}\right)^2.
\end{equation}
We note that the eigenstates $\phi_n(Z)$ are Landau polariton states shifted by the external electric field. 

\begin{figure}[H]
\begin{center}
  \includegraphics[height=5cm, width=\linewidth]{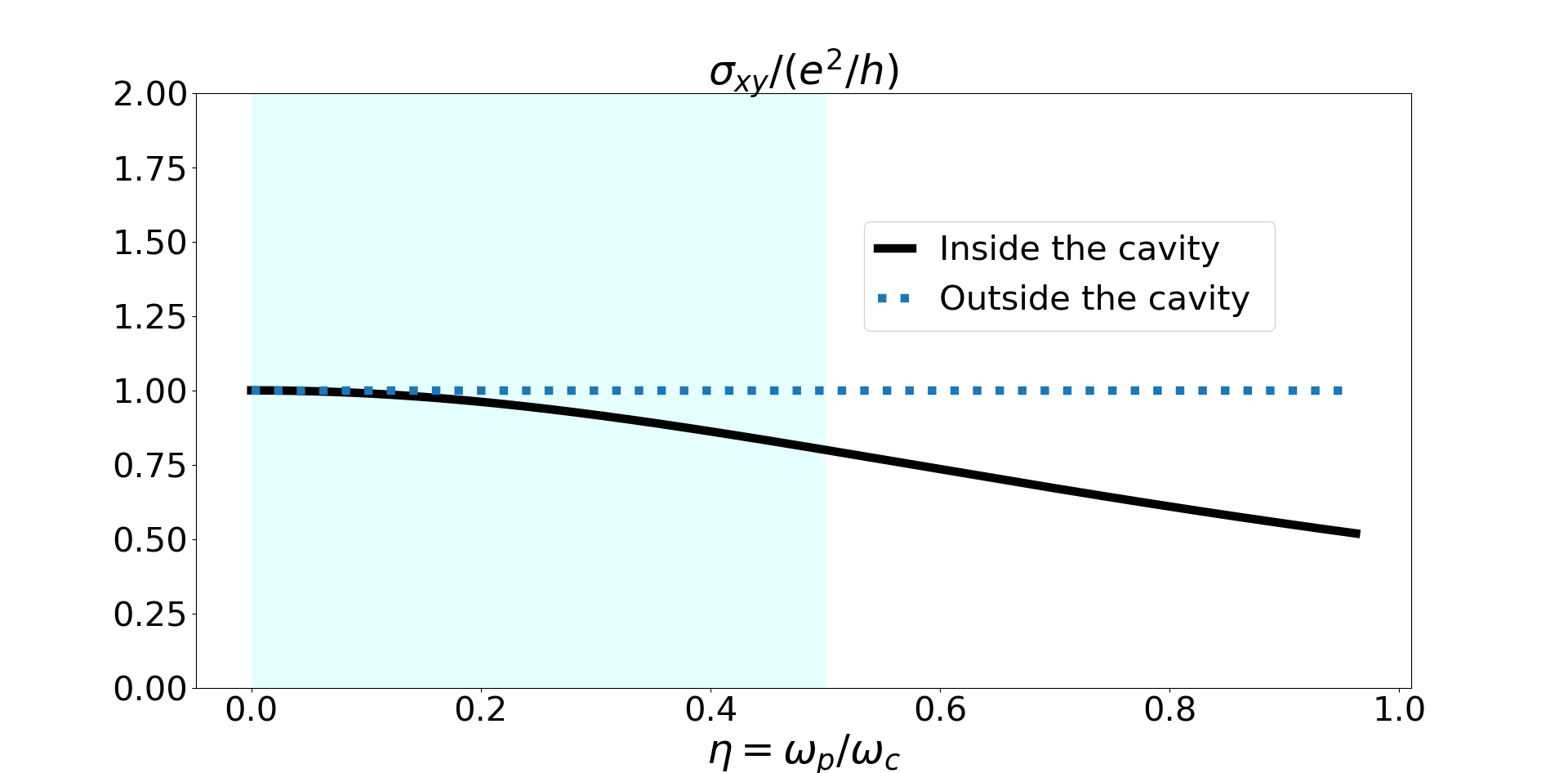}
\caption{\label{Hall Cavity g}Depiction of the first quantum Hall plateau as a function of the light-matter coupling constant $\eta$. For $\eta=0$ we obtain the standard value of the Hall conductance. As the light-matter coupling increases the Hall conductance decreases. This effect can be understood as a screening or renormalization effect due to the cavity. The shaded area indicates the regime in which typically experiments are performed.}   
\end{center}
\end{figure}

From the energy spectrum it is clear that the degeneracy with respect to $k_x$ is now lifted due to the electric field. But here the strength of the electric field $E$ is considered to be much smaller than the strength of the magnetic field $B$, which implies that $E/B \approx 0$. As a consequence, the degeneracy with respect to $k_x$ remains effectively the same. Further, having the expression for the energies we can straightforwardly compute the group velocity of the Landau polariton states in the $x$ direction~\cite{Mermin}
\begin{equation}
v_x=\frac{1}{\hbar}\frac{\partial E_{n,k_x}}{\partial k_x}=\frac{E}{B(1+\eta^2)}.
\end{equation}
The total current of the system in the $x$ direction $J_x$ is obtained by summing the velocity $v_x$ over all occupied Landau-polariton states,
\begin{equation}
J_x=e\sum^{\nu-1}_{n=0}\frac{L}{2\pi}\int^{\frac{eBL}{\hbar}}_0 v_x dk_x.
\end{equation}
Here we assumed that we have $\nu$ Landau-polariton states occupied, from $n=0$ to $n=\nu-1$. Further, in order to specify the region of integration for the electronic momentum $k_x$, we followed the standard procedure used in the Landau-level setting, where the system is considered to have a finite size in 2D with area $S=L^2$ and the momenta $k_x$ range from $0$ to $e BL/\hbar$~\cite{Tong, Peierls}. After integrating over $k_x$, summing over $n$, and dividing by the area of the system $S=L^2$, we find that the total current density $j_x=J_x/S$ induced by the electric field is
\begin{equation}
j_x=\frac{e^2\nu}{h(1+\eta^2)}E.
\end{equation}
From the above expression we deduce that the Hall conductance $\sigma_{xy}$ of the 2D Landau levels coupled to the cavity field is still quantized and is given by the expression 
\begin{equation}\label{Hall cavity}
\sigma_{xy}=\frac{e^2}{h(1+\eta^2)}\nu \;\;\; \textrm{with}\;\; \nu \in \mathbb{N}.
\end{equation}
The Hall conductance depends on the electron charge $e$, on Planck's constant $h$, on the amount of occupied Landau-polariton states $\nu$ and on the light-matter coupling between the Landau levels and the cavity photons $\eta=\omega_p/\omega_c$. This result makes clear that the cavity modifies the plateaus of the Hall conductance in the integer regime. This is a significant result as it demonstrates that the cavity field, due to its long-range nature, can circumvent the topological protection of edge states which carry the Hall current and modify the fundamental Hall conductance. In the limit of the light-matter coupling going to zero, $\eta\to 0$, we recover the standard result for the quantization of the Hall conductance $\sigma_{xy}=e^2\nu/h$ as it is shown in Fig.~\ref{Hall Cavity g}. This is a beautiful consistency check of our QED-Bloch theory. However, as the light-matter coupling increases, the value of the quantum Hall plateaus decrease as it is shown in Fig.~\ref{Hall Cavity g}. This decrease of the Hall plateaus can be understood as a renormalization or screening effect due to the strong vacuum fluctuations induced by the cavity field. 

To gain more understanding on the behavior of the Hall conductance inside the cavity we would also like to plot the cavity-modified Hall conductance given by Eq.~(\ref{Hall cavity}) as a function of the strength of the external magnetic field $B$. To do so, we rewrite the dimensionless light-matter coupling constant $\eta=\omega_p/\omega_c$ as a function of the magnetic field $B$ by simply using the definition for the cyclotron frequency, $\omega_c=eB/\me$. Upon this substitution $\eta$ takes the form
\begin{equation}
\eta=\frac{\me\omega_p}{e B}=\frac{B_{\textrm{cav}}}{B} \;\;\; \textrm{where}\;\;\; B_{\textrm{cav}}=\frac{\me\omega_p}{e}.
\end{equation}
We note that the quantity $B_{\textrm{cav}}$ has magnetic-field dimensions and it describes the magnetic field strength that corresponds to the diamagnetic frequency $\omega_p$ of the cavity. For $\omega_p$ at the order of one THz, as in the setting of Landau polaritons~\cite{paravacini2019, ScalariScience, li2018}, the strength of the cavity field is on the order of one Tesla.

\begin{figure}[H]
\begin{center}
  \includegraphics[height=5cm, width=\linewidth]{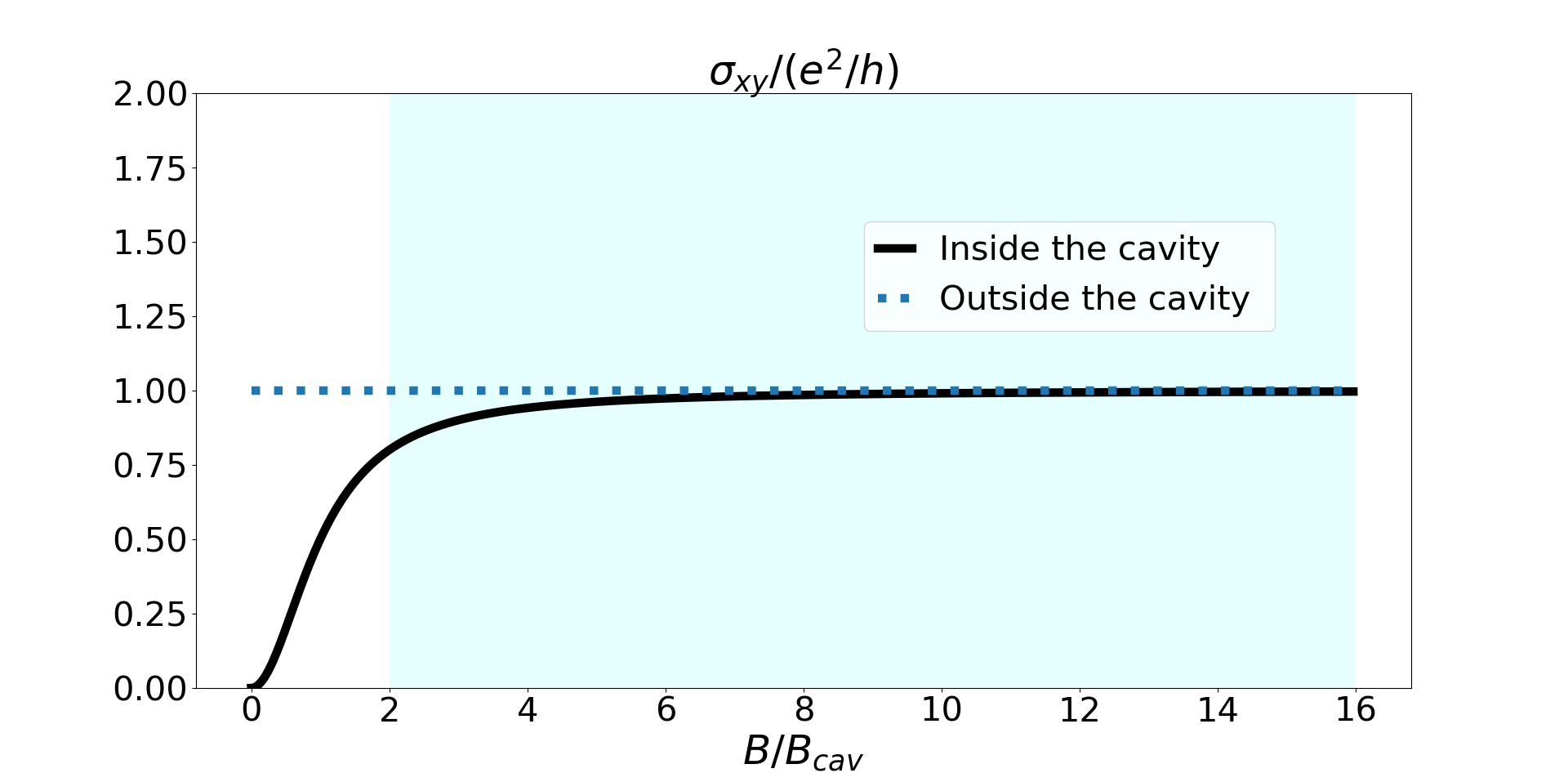}
\caption{\label{Hall Cavity B}First quantized Hall plateau as a function of the ratio between the strength of the external magnetic field and the strength of the cavity magnetic field $B/B_{\textrm{cav}}$. In the region where the cavity strength is larger, the Hall conductance deviates strongly from its standard value $e^2/h$. As the external magnetic field increases and becomes larger than the strength of the cavity we obtain the standard quantized Hall plateau. The shaded area indicates the regime in which typically experiments are performed.}   \end{center}
\end{figure}

Substituting now the expression for $\eta$ as a function of the magnetic field, the Hall conductance takes the form
\begin{equation}
\sigma_{xy}=\frac{e^2\nu}{h}\frac{1}{1+B^2_{\textrm{cav}}/B^2}.
\end{equation}
In Fig.~\ref{Hall Cavity B} we plot the Hall conductance as a function of the relative magnetic field $B/B_{\textrm{cav}}$. Figure~\ref{Hall Cavity B} shows that in the regime where the strength of the cavity field $B_{\textrm{cav}}$ is larger than the external magnetic field $B$, the cavity-modified Hall conductance deviates strongly from the value of the Hall plateau outside the cavity. As the strength of the external magnetic field increases and becomes larger than the cavity field, $B>B_{\textrm{cav}}$ (this is the common regime in experiments), the Hall conductance approaches the value outside the cavity and demonstrates the standard Hall plateau. This result provides an alternative understanding, from a QED point of view, of why strong magnetic fields are required for the formation of an exact Hall plateau. Namely, that the external magnetic field should reach such a value that it becomes stronger and dominates the vacuum fluctuations of the quantized electromagnetic field. 
Finally, we note that in the case where many modes contribute to the light-matter coupling, the strength of the cavity magnetic field gets enhanced and renormalizes the light-matter coupling and the Hall conductance as we show in Appendix~\ref{Many Modes}.

 \section{Summary \& Outlook}\label{Conclusions}

In this article we studied two-dimensional non-interacting quantum Hall systems strongly coupled to the quantized photon field of a cavity in the optical limit (or dipole approximation). To do so, we employed the recently proposed QED-Bloch theory in which the broken translational symmetry due to an external homogeneous magnetic field is restored by taking into account the quantum fluctuations of the photon field~\cite{rokaj2019} and which thus provides a first-principles framework for the description of periodic materials in homogeneous magnetic fields and strongly coupled to the photon field. We reviewed the basic steps in the construction of QED-Bloch theory and we introduced the single-particle effective Hamiltonian which has been successful in the description of Landau polaritons~\cite{rokaj2019}. In this general framework we constructed a QED-generalization of the Bloch ansatz (the QED-Bloch ansatz) and we derived the respective QED-Bloch central equation.

Subsequently, we applied the effective Hamiltonian to the study of two particular quantum Hall systems under strong coupling to the quantized cavity field: (i) the Hofstadter butterfly and (ii) the integer quantum Hall effect in the regime of non-interacting Landau levels.

\textbf{Polaritonic Hofstadter Butterfly}.---In the case of a non-interacting 2D periodic material under cavity confinement and in the presence of a homogeneous magnetic field we found that for the energy spectrum of the system as a function of the dimensionless light-matter coupling constant a self-similar pattern emerges which we call the polaritonic Hofstadter butterfly (see Fig.~\ref{Fractal_pol_g5}). This polaritonic fractal is an extension of the standard Hofstadter butterfly~\cite{Hofstadter} in cavity QED. With the advent of Moir\'{e} materials the Hofstadter butterfly has become now experimentally accessible through magnetotransport measurements of the Wannier diagram (integrated density of states)~\cite{Wannier, DeanButterfly, WangButterfly, BarrierButterfly, ForsytheButterfly}. We believe that our prediction of the existence of fractal polaritonic spectra due to strong light-matter coupling can be observed in such Moir\'{e} systems under cavity confinement via transport measurements. Further, the prediction of the polaritonic butterfly opens a new avenue for the exploration of fractal physics in the field of cavity QED. This novel polaritonic fractal could potentially be interesting also from a mathematical point of view, like the original Hofstadter butterfly~\cite{Hofstadter, Avila2006}, and might provide new connections between mathematics and physics~\cite{Langlands_Butterfly}.

\textbf{Cavity Engineering of the Hofstadter Butterfly.}---In addition, we computed the energy spectrum for a periodic material as a function of the relative magnetic flux and we found that for a terahertz cavity~\cite{ScalariScience, paravacini2019, li2018} the standard Hofstadter butterfly gets modified due to the strong vacuum fluctuations of the photon field (see Fig.~\ref{Hofstadter fluctuations large}). This phenomenon is most prominent in the intermediate regime of not exceedingly large magnetic fluxes, where the cavity field dominates. The modification of the Hofstadter butterfly should again be observable via transport measurements in the respective Wannier diagram~\cite{Wannier}. The Wannier diagram might provide a straightforward path to the observation of cavity effects on the Hofstadter spectrum. Finally, we also compared our results of the cavity engineering of the Hofstadter butterfly to the Floquet driving of the butterfly~\cite{FloquetButterflyKagome, FloquetButterflySquare, GenesisFloquetButterfly} and we found a basic consistency between QED-Bloch theory and Floquet theory.    

\textbf{Semi-classical Limit of QED-Bloch Theory.}---Further, we showed that our QED-Bloch theory and the QED-Bloch central equation (\ref{QED-Bloch Central}) are applicable also in the semi-classical limit of no quantized field. In this limit our theory recovers the standard Hofstadter butterfly fractal spectrum and provides a first-principles framework for its description. We believe it can thus help to understand recent experiments performed on Moir\'{e} systems~\cite{DeanButterfly, WangButterfly, BarrierButterfly, ForsytheButterfly}. Moreover, in the semi-classical limit, the dual relation between the minimal-coupling Hamiltonian and the
tight-binding models with the Peierls phase was described (see Fig.~\ref{Butterfly Duality}). This duality is of fundamental importance for understanding the behavior of periodic materials perpendicular to homogeneous magnetic fields and has been noted in experimental studies of the Hofstadter butterfly~\cite{DeanButterfly}.

\textbf{Modification of the Integer Hall Effect.}---As a further application of QED-Bloch theory we considered the quantum Hall effect in the integer regime when strongly coupled to the photon field of a cavity. In this case, our system consists of non-interacting 2D Landau levels coupled to the cavity. Due to the strong coupling between the cavity photons and the Landau levels hybrid quasiparticles emerge, known as Landau polaritons~\cite{rokaj2019,ScalariScience,paravacini2019}. The formation of the Landau polaritons modifies the plateaus of the Hall conductance, $\sigma_{xy}=e^2\nu/h(1+\eta^2)$, which now depend on the dimensionless light-matter coupling $\eta$. This is a very important result because it demonstrates that the long-range nature of the cavity photon-field circumvents the topological protection of the integer Hall effect and actually modifies this fundamental phenomenon of condensed matter physics. We believe that the modification of the Hall conductance can be measured for a 2DEG inside a cavity and could provide further insights on the recently observed modification of the integer Hall effect~\cite{FaistCavityHall}. In connection to these exciting, recent experiments, we would like to mention that the modification of the Hall plateaus in Ref.~\cite{FaistCavityHall} is attributed to the presence of impurities and disorder in the 2DEG. The cavity field benefits from the presence of impurities and mediates a hopping mechanism which breaks the topological protection of the Hall effect~\cite{CiutiHopping}. In the absence of impurities, however, the proposed cavity-mediated hopping vanishes. On the other hand, our work proposes a mechanism for the modification of the Hall plateaus in the absence of impurities, for a clean 2DEG, without any disorder. Our prediction should be interpreted as a renormalization effect (analogous to a Lamb shift~\cite{Lambshift}) of the Hall plateaus or equivalently as a screening effect on the external magnetic field by the internal magnetic field of the cavity (see Fig.~\ref{Hall Cavity B}). It is important to emphasize that the two approaches, i.e., the cavity-mediated hopping and our renormalization effect, are not contradictory but rather complementary as they apply to different physically relevant settings. In conclusion, our findings provide new insights and pave the way for the exploration of quantum Hall physics, in the integer and the fractional regime, embedded in the field of cavity QED.

\begin{acknowledgements}
We would like to thank J.~Faist for useful discussions. This work was supported by the European Research Council (ERC-2015-AdG694097), the Cluster of Excellence ``Advanced Imaging of Matter'' (AIM), Grupos Consolidados (IT1249-19), SFB925 ``Light induced dynamics and control of correlated quantum systems'', the Austrian Science Fund (J 4107-N27), and the Deutsche Forschungsgemeinschaft (DFG, German Research Foundation) via the Emmy Noether program (SE 2558/2). The Flatiron Institute is a division of the Simons Foundation.
\end{acknowledgements}

\appendix

\section{Displacement Operator Algebra}\label{Displacement Algebra}

The aim of this appendix is to show how the matrix elements defined in Eq.~(\ref{matrix elements}) can be computed and how the result in Eq.~(\ref{Matrixeq}) was obtained. The matrix elements that we are interested in computing are
\begin{equation}\label{matrix el app}
\langle \phi_i(v-A^{k_x}_n)|\rme^{-\textrm{i}G^v_{\bi{n}-\bi{n}^{\prime}}v}|\phi_j(v-A^{k_x}_{n^{\prime}})\rangle.
\end{equation}
To calculate these matrix elements we will first perform a change of coordinates, $s=v-A^{k_x}_n$, which will give us an overall phase independent of the integration, and we have for the matrix elements of Eq.~\eqref{matrix el app}
\begin{equation}
\rme^{-\textrm{i}A^{k_x}_nG^v_{\bi{n}-\bi{n}^{\prime}}}\langle \phi_i\left(s\right)|\rme^{-\textrm{i}G^v_{\bi{n}-\bi{n}^{\prime}}s}|\phi_j\left(s+A^0_{n-n^{\prime}}\right)\rangle.
\end{equation}
In order to compute the matrix elements above we will use the algebra of displacement operators~\cite{Glauber}. The plane wave $\exp(-\textrm{i}G^v_{\bi{n}-\bi{n}^{\prime}}s)$ can be written as a displacement operator by using the expression for the coordinate $s$ in terms of the annihilation and creation operators $\hat{b},\hat{b}^{\dagger}$~\cite{GriffithsQM},
\begin{equation}
s=\sqrt{\frac{\hbar}{2\mu\Omega}}\left(\hat{b}+\hat{b}^{\dagger}\right).
\end{equation}
Using the latter, we have for the plane wave in terms of the displacement operator
\begin{equation}
\rme^{-\textrm{i}G^v_{\bi{n}-\bi{n}^{\prime}}s}=\hat{D}\left(-\textrm{i}\sqrt{\frac{\hbar}{2\mu\Omega}} G^v_{\bi{n}-\bi{n}^{\prime}}\right).
\end{equation}
In addition, the wavefunction $\phi_j(s+A^0_{n-n^{\prime}})$ can be written as
\begin{equation}
\phi_j\left(s+A^0_{n-n^{\prime}}\right)=\hat{T}\left(A^0_{n-n^{\prime}}\right)\phi_j(s)
\end{equation}
using the translation operator, which is given by the expression~\cite{Mermin}
\begin{equation}
\hat{T}\left(A^0_{n-n^{\prime}}\right)=\exp\left(A^0_{n-n^{\prime}}\partial_s\right).
\end{equation}
The differential operator $\partial_s$ in terms of annihilation and creation operators is 
\begin{equation}
\partial_s\equiv\frac{\partial}{\partial s}=\sqrt{\frac{\mu\Omega}{2\hbar}}\left(\hat{b}-\hat{b}^{\dagger}\right).
\end{equation}
This implies that the translation operator can also be written as a displacement operator~\cite{Glauber},
\begin{equation}
\hat{T}\left(A^0_{n-n^{\prime}}\right)=\hat{D}\left(-\sqrt{\frac{\mu\Omega}{2\hbar}}A^0_{n-n^{\prime}}\right).
\end{equation}
Using the expressions we derived in terms of the displacement operators we obtain the following expression for the matrix elements in Eq.~(\ref{matrix el app}).
\begin{equation}\label{matrixeqAp}
\rme^{-\textrm{i}A^{k_x}_n G^v_{\bi{n}-\bi{n}^{\prime}}} \langle \phi_i|\hat{D}\left(-\frac{\textrm{i}\sqrt{\hbar}G^v_{\bi{n}-\bi{n}^{\prime}}}{\sqrt{2\mu\Omega}}\right)\hat{D}\left(-\sqrt{\frac{\mu\Omega}{2\hbar}}A^0_{n-n^{\prime}}\right)|\phi_j\rangle
\end{equation}
We now use the formula from Cahill-Glauber~\cite{Glauber}
\begin{equation}
\hat{D}(\alpha)\hat{D}(\beta)=\hat{D}(\alpha+\beta)\exp((\alpha\beta^*-\alpha^*\beta)/2)
\end{equation}
and we obtain the following result for the product of displacement operators
\begin{equation}
\begin{aligned}
&\hat{D}\left(-\textrm{i}\sqrt{\frac{\hbar}{2\mu\Omega}} G^v_{\bi{n}-\bi{n}^{\prime}}\right)\hat{D}\left(-\sqrt{\frac{\mu\Omega}{2\hbar}}A^0_{n-n^{\prime}}\right)\\
=&\hat{D}\left(\alpha_{\bi{n}-\bi{n}^{\prime}}\right)\rme^{\frac{\textrm{i}}{2}G^v_{\bi{n}-\bi{n}^{\prime}}A^0_{n-n^{\prime}}},
\end{aligned}
\end{equation}
where the matrix elements $\alpha_{\bi{n}-\bi{n}^{\prime}}$ are
\begin{equation}\label{alphamatrixAp}
\alpha_{\bi{n}-\bi{n}^{\prime}}=-\sqrt{\tfrac{\mu\Omega}{2\hbar}}A^0_{n-n^{\prime}}-\textrm{i}\sqrt{\tfrac{\hbar}{2\mu\Omega}}G^v_{\bi{n}-\bi{n}^{\prime}}.
\end{equation}
We substitute the expression for the product of the displacement operators into Eq.~(\ref{matrixeqAp}) and we have
\begin{equation}\label{Equation}
\rme^{-\textrm{i}G^v_{\bi{n}-\bi{n}^{\prime}}A^{k_x}_{(n+n^{\prime})/2}} \langle\phi_i|\hat{D}\left(\alpha_{\bi{n}-\bi{n}^{\prime}}\right)|\phi_j\rangle.
\end{equation}
The matrix representation of this displacement operator in the basis $\{\phi_i(s)\}$ is given by~\cite{Glauber}
\begin{eqnarray}\label{displacementeqAp}
\langle \phi_i|\hat{D}(\alpha_{\bi{n}-\bi{n}^{\prime}})|\phi_j\rangle=\sqrt{\frac{j!}{i!}}\alpha^{i-j}_{\bi{n}-\bi{n}^{\prime}}\rme^{-\frac{|\alpha_{\bi{n}-\bi{n}^{\prime}}|^2}{2}}L^{(i-j)}_j(|\alpha_{\bi{n}-\bi{n}^{\prime}}|^2) \nonumber\\
\end{eqnarray}
where $i\geq j$ and $L^{(i-j)}_j(|\alpha_{\bi{n}-\bi{n}^{\prime}}|^2)$ are the associated Laguerre polynomials. We note that for $j>i$ one needs to take 
\begin{equation}
\langle \phi_i|\hat{D}(\alpha_{\bi{n}-\bi{n}^{\prime}})|\phi_j\rangle=(-1)^{j-i}\langle \phi_j|\hat{D}(\alpha_{\bi{n}-\bi{n}^{\prime}})|\phi_i\rangle^{*}
\end{equation}
because $\hat{D}^{\dagger}(\alpha)=\hat{D}(-\alpha)$~\cite{Glauber}. Finally, combining the result that we obtained in Eq.~(\ref{Equation}) with the previous definitions, we obtain the expression for the matrix elements in Eq.~(\ref{matrix el app})
\begin{widetext}
\begin{equation}
\langle \phi_i(v-A^{k_x}_n)|\rme^{-\textrm{i}G^v_{\bi{n}-\bi{n}^{\prime}}v}|\phi_j(v-A^{k_x}_{n^{\prime}})\rangle=\rme^{-\textrm{i}G^v_{\bi{n}-\bi{n}^{\prime}}A^{k_x}_{(n+n^{\prime})/2}}\langle \phi_i|\hat{D}(\alpha_{\bi{n}-\bi{n}^{\prime}})|\phi_j\rangle.
\end{equation}
 \end{widetext}

\section{Semi-classical Limit or No-quantized-field Limit}\label{No quantized field}

In this appendix we are interested in performing the semi-classical limit of no quantized field for our QED-Bloch central equation~(\ref{QED-Bloch Central}). In this limit the 2D periodic material is only under the influence of the external magnetic field $\bi{B}_{\textrm{ext}}$, while the quantized field $\hat{\bi{\mathcal{A}}}$ goes to zero. Mathematically, this limit can be performed by taking the diamagnetic frequency to zero, $\omega_p\rightarrow 0$, because the quantized field $\hat{\bi{\mathcal{A}}}$ is proportional to $\omega_p$ (see Eq.~(\ref{effective vector potential})). This limit is equivalent to taking the light-matter coupling $\eta$ to zero, $\eta\rightarrow 0$. Our QED-Bloch central equation~\eqref{QED-Bloch Central} was
\begin{widetext}
\begin{equation}
U^{\bi{k}}_{\bi{n},i} \Bigg[\frac{\hbar^2(k_w+G^w_{\bi{n}})^2}{2M}+\mathcal{E}_i-E_{\bi{k}}\Bigg]+\sum_{\bi{n}^{\prime},j}V_{\bi{n}-\bi{n}^{\prime}} U^{\mathbf{k}}_{\mathbf{n}^{\prime},j}\; \rme^{-\textrm{i}G^v_{\bi{n}-\bi{n}^{\prime}}A^{k_x}_{(n+n^{\prime})/2}}\langle \phi_i|\hat{D}(\alpha_{\bi{n}-\bi{n}^{\prime}})|\phi_j\rangle=0
\end{equation}
\end{widetext}
and we will take the limit $\omega_p \rightarrow 0$ for all the parameters in the above equation. First, we consider the limit $\omega_p \rightarrow 0$ for the mass parameter $M$ defined in Eq.~(\ref{mass polaritonic parameters}) and we find that $\lim_{\omega_p\rightarrow 0}M = \infty$.
This implies that the kinetic term depending on $k_w$ in the central equation vanishes and the Fourier components of our polaritonic Bloch wave no longer depend on $k_w$. Due to the vanishing of the $w$ degree of freedom the index $m$ in the Bloch wave becomes redundant,
\begin{equation}
U^{k_x,k_w}_{n,m,j} \longrightarrow U^{k_x}_{n,j}.
\end{equation}
Consequently, the central equation reduces to
\begin{widetext}
\begin{equation}
U^{k_x}_{n,i} \left(\mathcal{E}_i-E_{k_x}\right)+\sum_{n^{\prime},m^{\prime},j}V_{n-n^{\prime},m^{\prime}} U^{k_x}_{n^{\prime},j}\; \rme^{-\textrm{i}G^v_{n-n^{\prime},m^{\prime}}A^{k_x}_{(n+n^{\prime})/2}}\langle \phi_i|\hat{D}(\alpha_{n-n^{\prime},m^{\prime}})|\phi_j\rangle=0.
\end{equation}
\end{widetext}
To obtain the result above we also relabelled the index $-m^{\prime} \rightarrow m^{\prime}$. Now what is left to be done is to perform the $\omega_p\rightarrow 0$ limit for the rest of the parameters in the central equation which depend on $\omega_p$. By doing so we find
\begin{eqnarray}\label{Omega-limit}
&&\lim_{\omega_p\rightarrow 0}\Omega =\omega_c,\;\; \lim_{\omega_p\rightarrow 0 }\mu\Omega=\tfrac{2\me}{\omega_c},\\
&&\lim_{\omega_p \rightarrow 0} G^v_{n-n^{\prime},m^{\prime}}=\tfrac{\sqrt{2}}{\omega_c}G_{n-n^{\prime},m^{\prime}}
\end{eqnarray}
and
\begin{equation}\label{beta matrix}
\begin{aligned}
\lim_{\omega_p\rightarrow 0}\alpha_{n-n^{\prime},m^{\prime}}&=\sqrt{\tfrac{\hbar}{2\me\omega_c}}\left(- G^x_{n-n^{\prime}}-\textrm{i}G_{n-n^{\prime},m^{\prime}}\right) \\
&\equiv \beta_{n-n^{\prime},m^{\prime}}.
\end{aligned}
\end{equation}
Substituting all the above results and the definition for $A^{k_x}_{(n+n^{\prime})/2}$ given by Eq.~(\ref{HO-shift}) we have
\begin{widetext}
\begin{equation}\label{LLB Central}
U^{k_x}_{n,i}\left[\hbar\omega_c\left(i+\frac{1}{2}\right)-E_{k_x}\right]+\sum_{n^{\prime},m^{\prime},j}V_{n-n^{\prime},m^{\prime}} U^{k_x}_{n^{\prime},j} \rme^{\frac{-\textrm{i}\hbar(k_x+\frac{1}{2}G^x_{n+n^{\prime}})G_{n-n^{\prime},m^{\prime}}}{\me\omega_c}}\langle \phi_i|\hat{D}(\beta_{n-n^{\prime},m^{\prime}})|\phi_j\rangle=0.
\end{equation}
\end{widetext}
The central equation derived depends solely on electronic parameters like the electronic crystal momentum $k_x$, the mass of the electron $\me$ and the cyclotron frequency $\omega_c=eB/\me$, which is characteristic for electrons in a constant magnetic field~\cite{Landau}. As a consequence, the above central equation describes consistently the physics of two-dimensional periodic systems in the presence of a perpendicular homogeneous magnetic field. From this equation we can compute the energy bands for such systems for all values of the magnetic field because our approach is non-perturbative and does not rely on the magnetic translation group which puts particular restrictions on the value of the magnetic field~\cite{BrownMTG, MTG_I, MTG_II}.   

For completeness, we would also like to give the expression of the polaritonic Bloch ansatz defined in Eq.~(\ref{BlochAnsatzHermite}) in the limit of no quantized field. The QED-Bloch ansatz depends on the polaritonic coordinates $w$ and $v$ defined in Eq.~(\ref{w and v coordinates}). Taking the no-quantized-field limit $\omega_p\rightarrow 0$ the coordinate $w$ vanishes while the coordinate $v$ becomes $v=-\omega_c y$. Thus, we find that the polaritonic QED-Bloch ansatz in the limit of no quantized field is 
\begin{equation}\label{LLB ansatz}
\Psi_{k_x}(x,y)=\rme^{\textrm{i}k_x}\sum_{n,j}U^{k_x}_{n,j}\rme^{\textrm{i}G^x_n}\phi_j\left(-\frac{\omega_cy}{\sqrt{2}}-A^{k_x}_n\right).
\end{equation}
The above wavefunction corresponds to a correlated expansion between Bloch waves in the $x$ coordinate and Landau levels $\phi_j(-\omega_cy/\sqrt{2}-A^{k_x}_n )$ in the $y$ coordinate. Such an expansion has been used for the description of 2D materials in homogeneous magnetic fields in several publications~\cite{PfannkucheButterfly, Langbein1969, Rauh1975, GeiselButterflyChaos} and central equations analogous to Eq.~(\ref{LLB Central}) have been derived.   

\section{The Effect of Many Modes}\label{Many Modes}

In this appendix we will look into the effect of many modes for the strength of the light-matter coupling and the corresponding implication for the Hall conductance. To do so, we will not try to take into account exactly a finite amount of modes, but we will rather follow an effective approach in which the dependence of the single-mode coupling constant $\eta$ on the photonic momenta is introduced back and then the sum of the single-mode couplings over the photonic momenta defines an effective many-mode coupling. This effective approach was also followed in Ref.~\cite{rokaj2020} and it was shown that it captures the exact running of the light-matter coupling as a function of the photonic upper cutoff and recovers well-known perturbative mass-renormalization results of quantum field theory.

The single-mode coupling $\eta$ depends on the cavity frequency $\omega_{\textrm{cav}}$ via the diamagnetic frequency $\omega_p$ defined in Eq.~(\ref{plasma frequency}). The cavity frequency itself is a function of the photonic momentum $\kappa_z=\pi n_z/L_z$, $\omega_{\textrm{cav}}=c|\kappa_z|$, where we take only the out-of-plane momenta into account. Substituting this expression for $\omega_{\textrm{cav}}$ into the definition of $\eta$ we find $\eta^2=e^2n_{\textrm{2D}}|\kappa_z|/\me\pi\epsilon_0\omega^2_c$. Then the effective many-mode light-matter coupling constant is 
\begin{equation}
\eta^2(\Lambda_0)=\frac{e^2n_{\textrm{2D}}}{\me\pi \epsilon_0\omega^2_c}\frac{\pi}{L_z}\sum^{\Lambda_0}_{n_z=-\Lambda_0}|n_z|=\frac{\omega^2_p}{\omega^2_c}\Lambda_0(\Lambda_0+1)
\end{equation}
where the photonic momenta $\kappa_z$ were summed up to the upper cutoff $\Lambda=\Lambda_0\pi/L_z$, with $\Lambda_0 \in \mathbb{N}$, which is defined as an arbitrary multiple of the inverse of the cavity length $L_z$. For large $\Lambda_0\gg 1$ we can approximately consider $\Lambda_0+1\approx \Lambda_0$ and the many-mode effective coupling takes the simple form
\begin{equation}
\eta^2(\Lambda_0)=\Lambda^2_0\frac{\omega^2_p}{\omega^2_c}.
\end{equation}
The many-mode effective coupling $\eta(\Lambda_0)$ can be rewritten as the the ratio between the strength of the external magnetic field $B$ and the many-mode cavity magnetic field $B(\Lambda_0)=\Lambda_0\me\omega_p/e$ as $\eta(\Lambda_0)=B(\Lambda_0)/B$. We note that the many-mode cavity field $B(\Lambda_0)$ is merely a multiple of the of the single-mode cavity magnetic field in the singe-mode case $B_{\textrm{cav}}=\me\omega_p/e$. This means that the inclusion of many modes acts as an amplifier for the strength of the cavity field. Finally, replacing the single-mode coupling $\eta$ in the formula for the cavity-modified Hall conductance in Eq.~(\ref{Hall cavity}) with the effective many-mode coupling $\eta(\Lambda_0)$ one straightforwardly obtains the effect of many modes for the modified Hall conductance.

\bibliography{Fractal_polaritons}

\end{document}